\pdfoutput=1
\documentclass[11pt,twoside,a4paper,cmspaper,final,collab]{cms-tdr}
\def\svnVersion{ce1b78a}\def\svnDate{2025/01/02}\def\cmsCernNoTag{CERN-EP-2024-199}\def\cmsCernDate{\today}\def\cmsMessage{Published in the Journal of High Energy Physics as \href{http://dx.doi.org/10.1007/JHEP12(2024)172}{\doi{10.1007/JHEP12(2024)172}.}}
\begin{document}\cmsNoteHeader{HIG-21-004}

\cmsNoteHeader{HIG-21-004}

\title{Model-independent search for pair production of new bosons decaying into muons in proton-proton collisions at $\sqrt{s} = 13\TeV$}

\newlength\cmsTabSkip\setlength{\cmsTabSkip}{1ex}

\renewcommand{\Pa}{\textit{a}\xspace}

\newcommand{\gammaDark}{\ensuremath{{\PGg}_{\mathrm{D}}}\xspace}
\newcommand{\sDark}{\ensuremath{s_{\mathrm{D}}}\xspace}
\newcommand{\AsDark}{\ensuremath{\overline{s}_\mathrm{D}}\xspace}
\newcommand{\ZDark}{\ensuremath{Z_{\mathrm{D}}}\xspace}
\newcommand{\nOne}{\ensuremath{\textnormal{n}_{1}}\xspace}
\newcommand{\nDark}{\ensuremath{\textnormal{n}_{\mathrm{D}}}\xspace}
\newcommand{\PhOne}{\ensuremath{\Ph_{1}}\xspace}
\newcommand{\PhTwo}{\ensuremath{\Ph_{2}}\xspace}
\newcommand{\PhOneTwo}{\ensuremath{\Ph_{1,2}}\xspace}
\newcommand{\PaOne}{\ensuremath{\Pa_{1}}\xspace}
\newcommand{\PhI}{\ensuremath{\Ph_{i}}\xspace}
\newcommand{\dimuon}{\ensuremath{\PGm\PGm}\xspace}
\newcommand{\dimuonOne}{\ensuremath{\PGm\PGm_1}\xspace}
\newcommand{\dimuonTwo}{\ensuremath{\PGm\PGm_2}\xspace}
\newcommand{\MgammaDark}{\ensuremath{m_{{\PGg}_{\mathrm{D}}}}\xspace}
\newcommand{\MnOne}{\ensuremath{m_{{\textnormal{n}}_{1}}}\xspace}
\newcommand{\MnDark}{\ensuremath{m_{{\textnormal{n}}_{\mathrm{D}}}}\xspace}
\newcommand{\MsDark}{\ensuremath{m_{{\textnormal{s}}_{\mathrm{D}}}}\xspace}
\newcommand{\MZDark}{\ensuremath{m_{{\textnormal{Z}}_{\mathrm{D}}}}\xspace}
\newcommand{\MPhOne}{\ensuremath{m_{\Ph_{1}}}\xspace}
\newcommand{\MPhTwo}{\ensuremath{m_{\Ph_{2}}}\xspace}
\newcommand{\MPhOneTwo}{\ensuremath{m_{\Ph_{1,2}}}\xspace}
\newcommand{\MPhI}{\ensuremath{m_{\Ph_{i}}}\xspace}
\newcommand{\MPa}{\ensuremath{m_{\Pa}}\xspace}
\newcommand{\MPaOne}{\ensuremath{m_{\Pa_{1}}}\xspace}
\newcommand{\Mdimuon}{\ensuremath{m_{\PGm\PGm}}\xspace}
\newcommand{\MdimuonOne}{\ensuremath{m_{\PGm\PGm_1}}\xspace}
\newcommand{\MdimuonTwo}{\ensuremath{m_{\PGm\PGm_2}}\xspace}
\newcommand{\Mmuon}{\ensuremath{m_{\PGm}}\xspace}
\newcommand{\Mtau}{\ensuremath{m_{\PGt}}\xspace}
\newcommand{\TgammaDark}{\ensuremath{\tau_{{\PGg}_{\mathrm{D}}}}\xspace}
\newcommand{\cTgammaDark}{\ensuremath{c\tau_{{\PGg}_{\mathrm{D}}}}\xspace}
\newcommand{\CahLambda}{\ensuremath{C^{\text{eff}}_{\Pa\Ph}/\Lambda^{2}}\xspace}
\newcommand{\cll}{\ensuremath{C^{\text{eff}}_{ll}}\xspace}
\newcommand{\Zdimuon}{\ensuremath{z_{\dimuon}}\xspace}
\newcommand{\ZdimuonOne}{\ensuremath{z_{\dimuonOne}}\xspace}
\newcommand{\ZdimuonTwo}{\ensuremath{z_{\dimuonTwo}}\xspace}
\newcommand{\Ztrack}{\ensuremath{z_\text{track}}\xspace}
\newcommand{\Lxy}{\ensuremath{L_{xy}}\xspace}
\newcommand{\Lz}{\ensuremath{L_{z}}\xspace}
\newcommand{\DeltaY}{\ensuremath{\Delta y}\xspace}
\newcommand{\alphaGen}{\ensuremath{\alpha_\text{Gen}}\xspace}
\newcommand{\epsilonFull}{\ensuremath{\epsilon_{\text{Full}}}\xspace}
\newcommand{\SOne}{\ensuremath{S_\mathrm{I}(\Mdimuon)}\xspace}
\newcommand{\STwo}{\ensuremath{S_\mathrm{II}(\Mdimuon)}\xspace}
\newcommand{\SOneFinal}{\ensuremath{S_\mathrm{I}(\MdimuonOne)}\xspace}
\newcommand{\STwoFinal}{\ensuremath{S_\mathrm{II}(\MdimuonTwo)}\xspace}
\newcommand{\ASR}{\ensuremath{A_{\mathrm{SR}}}\xspace}
\newcommand{\ACR}{\ensuremath{A_{\mathrm{CR}}}\xspace}
\newcommand{\fSPS}{\ensuremath{f_\mathrm{SPS}}\xspace}
\newcommand{\muF}{\ensuremath{\mu_{\text{F}}}\xspace}
\newcommand{\muR}{\ensuremath{\mu_{\text{R}}}\xspace}
\newcommand{\functionMgammadark}{\ensuremath{f(\MgammaDark)}\xspace}
\newcommand{\PP}{\ensuremath{\Pp\Pp}\xspace}
\newcommand{\PhToTwoNOne}{\ensuremath{\Ph\to2\nOne}\xspace}
\newcommand{\PhOneTwoToPaOne}{\ensuremath{\PhOneTwo\to2\PaOne}\xspace}
\newcommand{\BrPhToGammaDarkX}{\ensuremath{\mathcal{B}(\Ph\to2\gammaDark+\PX)}\xspace}
\newcommand{\MassRangeNewLightBoson}{\ensuremath{0.25<\MPa<8.5\GeV}\xspace}
\newcommand{\NOneToDark}{\ensuremath{\nOne\to\nDark+\gammaDark}\xspace}
\newcommand{\EpsilonOverAlpha}{\ensuremath{\epsilonFull/\alphaGen}\xspace}
\newcommand{\BrPaOnetoPGm}{\ensuremath{\mathcal{B}(\PaOne\to2\PGm)}\xspace}
\newcommand{\PaOneToGluons}{\ensuremath{\PaOne\to\Pg\Pg}\xspace}
\newcommand{\Pmumu}{\ensuremath{P_{\PGm\PGm}}\xspace}
\newcommand{\Nsa}{\ensuremath{N_{\mathrm{SA}}}\xspace}
\newcommand{\Mmumui}{\ensuremath{m_{(\PGm\PGm)_{i}}}\xspace}
\newcommand{\Pggf}{\ensuremath{\Pg\Pg}\xspace}
\newcommand{\SignalRegionBackgroundTotal}{\ensuremath{7.95\pm1.12\stat\pm1.45\syst}\xspace}
\newcommand{\ControlRegionbbbarEvents}{\ensuremath{43}\xspace}
\newcommand{\SignalRegionbbbarEvents}{\ensuremath{7.26\pm1.11\stat}\xspace}
\newcommand{\SignalRegionObservedEvents}{\ensuremath{9}\xspace}
\newcommand{\ctau}{\ensuremath{c\tau}\xspace}

\date{\today}
\abstract{
The results of a model-independent search for the pair production of new bosons within a mass range of $0.21 < m < 60\GeV$, are presented. This study utilizes events with a four-muon final state. We use two data sets, comprising 41.5\fbinv and 59.7\fbinv of proton-proton collisions at $\sqrt{s} = 13\TeV$, recorded in 2017 and 2018 by the CMS experiment at the CERN LHC. The study of the 2018 data set includes a search for displaced signatures of a new boson within the proper decay length range of $0 < \ctau < 100\mm$. Our results are combined with a previous CMS result, based on 35.9\fbinv of proton-proton collisions at $\sqrt{s} = 13\TeV$ collected in 2016. No significant deviation from the expected background is observed. Results are presented in terms of a model-independent upper limit on the product of cross section, branching fraction, and acceptance. The findings are interpreted across various benchmark models, such as an axion-like particle model, a vector portal model, the next-to-minimal supersymmetric standard model, and a dark supersymmetric scenario, including those predicting a non-negligible proper decay length of the new boson. In all considered scenarios, substantial portions of the parameter space are excluded, expanding upon prior results.}

\hypersetup{
pdfauthor={CMS Collaboration},
pdftitle={Model-independent search for pair production of new bosons decaying into muons in proton-proton collisions at sqrt(s) = 13 TeV},
pdfsubject={CMS},
pdfkeywords={CMS, Higgs, ALP, dark photon, supersymmetry, vector portal}} 

\maketitle

\section{Introduction}\label{sec:intro}
Although the standard model (SM)~\cite{Englert:1964et, Higgs:1964pj, Guralnik:1964eu} of particle physics provides a multitude of high-precision predictions consistent with decades of experimental results, it does not explain the existence of dark matter~\cite{Trimble:1987ee, Bertone:2016nfn}. Many models have been proposed that predict new bosons as dark matter candidates~\cite{Maniatis:2009re, Duffy:2009ig, Arkani-Hamed:2008hhe}. 
To date, no direct experimental evidence has materialized for particles beyond the SM (BSM), and in that context the idea of new bosons with non-negligible proper decay length~\cite{Lee:2018pag} is of increasing interest.
In this paper, we present a model-independent search for the pair production of a new boson that decays into a pair of oppositely charged muons. This can happen in proton-proton (\PP) collisions as ${\PP\to2\Pa+\PX\to4\PGm+\PX}$, where \Pa is the new neutral boson and \PX are possible spectator particles~\cite{Belyaev:2010ka}. The new boson, \Pa, can be produced via various ``portals'' connecting to a hidden sector, such as a Higgs boson portal (either SM or non-SM) or a vector boson portal \cite{Arkani-Hamed:2008hhe, Rahmani_thesis, Curtin:2014cca}. Here, the production vertices of dimuons, \ie, pairs of oppositely-charged muons in the final state, can be either prompt or displaced. The displaced case results from the decays of long-lived BSM bosons. This generic signature enables us to set a model-independent limit on the product of the new boson production cross section, branching fraction to muons squared, and acceptance. With this model-independent limit, further interpretation is possible in models with four muons in the final state.

\begin{figure}[!htp]
\centering
\includegraphics[width=0.8\textwidth]{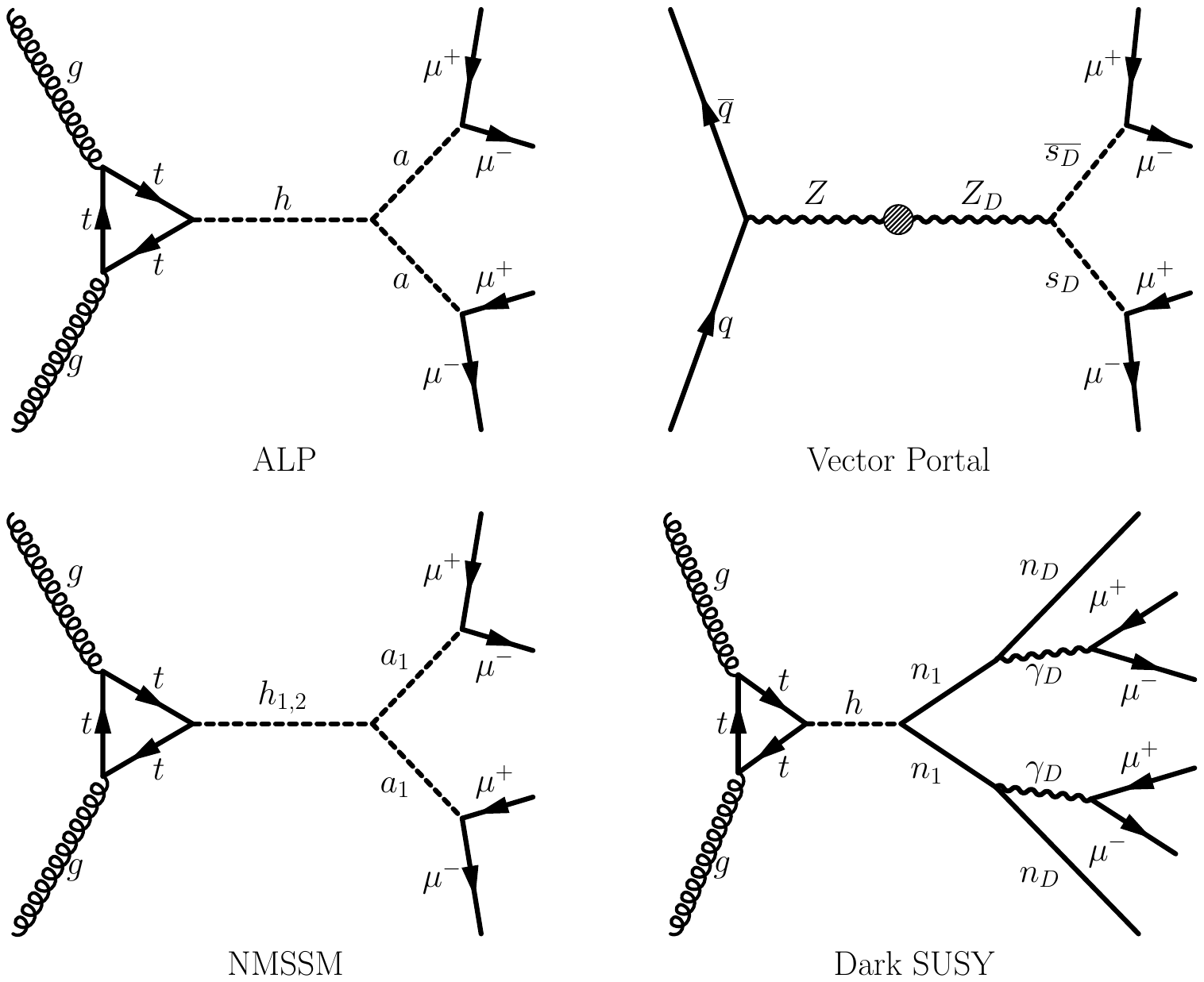}
\caption{\label{fig:feynmanDiagram} The Feynman diagrams of the benchmark signal models in the $s$-channel. Time moves from left to right.}
\end{figure}

We interpret the model-independent results in the context of several BSM benchmarks, including an axion-like particle (ALP) model~\cite{Armengaud:2013rta, Caputo:2024oqc, Bauer:2017ris, Duffy:2009ig, Weinberg:1977ma}, a vector portal model with a dark scalar boson \sDark~\cite{Buckley:2014fba, Wells:2008xg, Aguilar-Saavedra:2019iil, Abercrombie:2015wmb, Bishara:2016kjn, Baumgart:2009tn}, the next-to-minimal supersymmetric standard model (NMSSM)~\cite{Fayet:1974pd, Kaul:1981hi, Barbieri:1982eh, Nilles:1982dy, Frere:1983ag, Derendinger:1983bz, Drees:1988fc, Maniatis:2009re, Ellwanger:2009dp}, and supersymmetry (SUSY) models with hidden sectors (dark SUSY)~\cite{Arkani-Hamed:2008hhe, Baumgart:2009tn, Falkowski:2010cm}. In the ALP model~\cite{Bauer:2017ris}, the SM-like Higgs boson, \Ph, decays to the ALP, \Pa, via $\Ph \to 2\Pa$. The ALP then promptly decays into a dimuon. In the vector portal model, a massive dark vector boson \ZDark decays to two new scalar bosons \sDark, via $\ZDark \to \sDark \AsDark$, where we assume that the \sDark is not self-conjugate, \ie, it is not equal to its antiparticle. The dark scalar boson \sDark promptly decays to a dimuon. In the NMSSM, two of the three charge parity (CP) even, neutral Higgs bosons \PhOne or \PhTwo (generically denoted \PhOneTwo) can decay to one of the two CP-odd neutral Higgs bosons \PaOne via $\PhOneTwo \to 2\PaOne$. The CP-odd boson \PaOne promptly decays to a dimuon. In the dark SUSY scenario, the breaking of a new $U(1)_{\mathrm{D}}$ symmetry gives rise to a massive dark photon \gammaDark. This dark photon can couple to SM photons via a small kinetic mixing parameter, $\varepsilon$. The proper decay length of the dark photon depends on its mass, \MgammaDark, and $\varepsilon$. We use a signal topology where an SM-like \Ph decays to the lightest non-dark neutralino, \nOne, via $\Ph \to 2\nOne$. These neutralinos then decay via $\nOne \to \nDark + \gammaDark$, where \nDark is an undetectable dark neutralino. The dark photon \gammaDark then decays to a dimuon. Figure~\ref{fig:feynmanDiagram} displays the Feynman diagrams of the benchmark signal models.

The search presented in this analysis improves upon the previous result published by the CMS Collaboration in Ref.~\cite{CMS:2018jid}. Compared to this study, we add more data and add two new benchmark models, the ALP and vector portal models, and probe a larger mass range of the new, neutral boson \Pa ($0.21 < \MPa < 60\GeV$, compared to $0.25 < \MPa < 8.5\GeV$ in Ref.~\cite{CMS:2018jid}). Other searches at the LHC for ${\Ph \to 2\Pa}$ include the 4\PGm~\cite{CMS:2011xlr, CMS:2012qms, CMS:2015nay, CMS:2018jid}, 4\PGt~\cite{CMS:2015twz, CMS:2019spf}, 4\Pell~\cite{CMS:2021pcy, ATLAS:2015hpr, ATLAS:2021ldb, ATLAS:2024zxk, ATLAS:2021wob}, 4\PQb~\cite{ATLAS:2020ahi, ATLAS:2018pvw}, 4\PGg~\cite{ATLAS:2023ian}, 2\PQb2\PGt~\cite{CMS:2018zvv, CMS:2024uru}, 2\PGm2\PGt~\cite{CMS:2018qvj, CMS:2020ffa}, 6\PQq~\cite{LHCb:2016buh}, and collimated fermions~\cite{ATLAS:2023cjw, ATLAS:2022izj} final states. This model-independent analysis considers both promptly-decaying and long-lived ($\ctau < 100\mm$) bosons, and it is complementary to those studies exploring $\varepsilon$ in models with promptly-decaying~\cite{CMS:2019buh,CMS:2023hwl} and long-lived~\cite{CMS:2021sch,CMS:2022qej} bosons that decay to a dimuon.

Two data sets are used for this analysis, which include data collected using multiple triggers in 2017 and 2018. These data sets correspond to integrated luminosities of 41.5 and 59.7\fbinv of $\PP$ collisions at a center-of-mass energy of $\sqrt{s} = 13\TeV$, respectively. The results of the 2017 and 2018 data sets are first combined, corresponding to a total integrated luminosity of 101\fbinv. For the search mass range below 9\GeV, we also combine these two data sets with a previously published analysis that examined the 2016 data-taking period~\cite{CMS:2018jid}, leading to results in the search mass range up to 9\GeV that correspond to a combined integrated luminosity of 137\fbinv. Due to triggers using a primary vertex constraint in 2017, only 2018 data are used for the displaced signatures searches.

The CMS detector is improved in this data run with a new silicon pixel tracking detector~\cite{CMS:2012sda} installed and commissioned in 2017~\cite{Vami:2019gbk}. It features a larger detector volume with four layers in the barrel and three layers in the endcaps. Compared to the previous pixel detector, the innermost layer is positioned closer to the interaction point (IP) and the outermost layer is located farther from the IP. The new pixel detector can cope with a higher particle rate, improves impact parameter resolution, and increases the tracking efficiency. The performance of the hardware trigger algorithms for muons is also improved~\cite{CMS:2020cmk}. The analysis criteria are modified to accommodate the search in an extended model parameter space as compared to Ref.~\cite{CMS:2018jid}. New benchmark models are added to further test the model independence and diversify the interpretation.

This paper is organized as follows: First, we discuss the layout and operation of the CMS detector in Section~\ref{sec:cms}. Next, we review the simulation of signal samples in Section~\ref{sec:signal_modeling}, event selection criteria in Section~\ref{sec:data_selection}, and signal shape modeling in Section~\ref{sec:signal_shape}. We then present the processes of background estimation in Section~\ref{sec:bkg}, followed by a discussion of the systematic uncertainties in Section~\ref{sec:syst}. Finally, we provide the results of this analysis in Section~\ref{sec:results}, which is followed by a brief summary in Section~\ref{sec:summary}. Tabulated results are provided in the HEPData record for this analysis~\cite{hepdata}.

\section{The CMS detector}\label{sec:cms}
The central feature of the CMS apparatus is a superconducting solenoid of 6\unit{m} internal diameter, providing a magnetic field of 3.8\unit{T}. Within the solenoid volume are a silicon pixel and strip tracker, a lead tungstate crystal electromagnetic calorimeter (ECAL), and a brass and scintillator hadron calorimeter (HCAL), each composed of a barrel and two endcap sections. Forward calorimeters extend the pseudorapidity coverage provided by the barrel and endcap detectors. Muons are reconstructed in gas-ionization detectors embedded in the steel flux-return yoke outside the solenoid. 

Muons are measured in the pseudorapidity range $\abs{\eta} < 2.4$, with detection planes made using three technologies: drift tubes, cathode strip chambers, and resistive plate chambers. The single muon trigger efficiency exceeds 90\% over the full $\eta$ range, and the efficiency to reconstruct and identify muons is greater than 96\%. Matching muons to tracks measured in the silicon tracker results in a relative transverse momentum (\pt) resolution of 1\% in the barrel and 3\% in the endcaps, for muons with \pt up to 100\GeV. The \pt resolution in the barrel is better than 7\% for muons with \pt up to 1\TeV~\cite{CMS:2018rym}.

Events of interest are selected using a two-tiered trigger system~\cite{Khachatryan:2016bia}. The first level (L1), is composed of custom hardware processors, and uses information from the calorimeters and muon detectors to select events at a rate of around 100\unit{kHz} within a fixed time interval of about 4\mus. The second level, known as the high-level trigger (HLT), consists of a farm of processors running a version of the full event reconstruction software optimized for fast processing. The HLT further reduces the event rate to around 1\unit{kHz} before the data are sent for permanent storage~\cite{Khachatryan:2016bia}.

A more detailed description of the CMS detector, together with a definition of the coordinate system used and the relevant kinematic variables, can be found in Ref.~\cite{CMS:2008xjf}.

\section{Benchmark signal models}\label{sec:signal_modeling}
Simulated signal events are generated with Monte Carlo (MC) simulations and are used to determine the effect of the data selection criteria (described in Section~\ref{sec:data_selection}) on the various signal models. In the ALP model, production of the SM-like h via gluon-gluon (\Pggf) fusion and its subsequent decays are simulated at next-to-leading order (NLO) with the matrix-element generator \MGvATNLO2.4.2~\cite{Alwall:2014hca, Artoisenet:2012st}.
The mass of the SM-like h is fixed to 125\GeV and masses between 0.5 and 30\GeV for the ALP are simulated. For the vector portal model, production of the spin-1 \ZDark and its subsequent decays are simulated at NLO with \MGvATNLO2.6.5. Masses of the \ZDark between 85 and 200\GeV are simulated, thereby including the maximum \ZDark production cross section~\cite{Rahmani_thesis}; the masses of the scalar boson \sDark are simulated from 5 to 55\GeV depending on the kinematic constraints set by the \ZDark mass.

For the NMSSM, production of the CP-even \PhOneTwo via \Pggf fusion and its decay to the CP-odd \PaOne is carried out at leading order with \PYTHIA8.230~\cite{Sjostrand:2014zea}. Since the \PhOneTwo might not be the observed SM Higgs boson~\cite{ATLAS:2012yve, CMS:2012qbp, CMS:2013btf}, mass values of \MPhOneTwo between 90 and 150\GeV are simulated. This range is motivated by constraints set by the relic density measurements from WMAP~\cite{Hinshaw:2012aka}, Planck~\cite{Planck:2013pxb}, and searches at LEP~\cite{OPAL:2002ifx, OPAL:2002xgi, OPAL:2004vmm, DELPHI:2004bco, ALEPH:2010iye, ALEPH:2006tnd}. The mass of the boson \PaOne is set to vary between 0.5 and 3\GeV as motivated in Ref.~\citep{Dermisek:2010mg}.

In the dark SUSY model, production of the SM-like Higgs boson via \Pggf fusion and its decays are simulated at NLO with \mbox{\MGvATNLO2.4.2}. The masses of the SM-like Higgs boson, the neutralino \nOne, and the dark neutralino \nDark are fixed to 125, 60, and 1\GeV, respectively. Dark photon masses \MgammaDark are simulated between 0.25 and 58\GeV, and only the decays to oppositely charged muons are considered. Since the dark photon interacts weakly with SM particles, its decay width is negligible compared to the dimuon mass resolution.

Discussed in greater detail in Section~\ref{sec:data_selection}, we consider displaced tracks and vertices for the case of long-lived mediators in the 2018 analysis only. This is because the 2017 data set lacks triggers that are comparable to those used in the 2018 analysis for displaced signatures. When considering muon displacement for the 2018 analysis, we model the displaced vertices in the lab frame via an exponential distribution with \cTgammaDark between 0 and 100\mm. This proper decay length of 0 to 100\mm is required to achieve a uniform distribution of \EpsilonOverAlpha (see Sec.~\ref{sec:data_selection}) and to ensure that the displaced signal decays within the pixel detector volume so that at least one pixel hit can be required in the dimuon selection.

For all benchmark samples, \PP collisions at 13\TeV are simulated using a set of parton distribution functions (PDFs) provided by NNPDF3.1~\cite{NNPDF:2017mvq}. The parton shower, underlying event activity, and hadronization processes at the LHC are modeled with the MC event generator \PYTHIA8.230 using the CP5 tune~\cite{CMS:2019csb}. Pileup (PU), or additional collisions, are added to these events, which are run through the full CMS detector simulation based on \GEANTfour~\cite{GEANT4:2002zbu}. These events are then reconstructed with the same algorithms that are used for the data.

\section{Event selection}\label{sec:data_selection}
Several triggers were used to collect the data used in this study, all exhibiting a total efficiency that exceeds 90\% for all signal benchmarks described in Section~\ref{sec:signal_modeling}. For the 2017 data set, we employ a double muon trigger with \pt thresholds of 23 and 12\GeV. Three additional triggers with lower \pt thresholds are implemented to further improve the trigger efficiency for potential signals: a double muon trigger requiring muons of the same sign, with \pt thresholds of 18 and 9\GeV, a triple muon trigger with \pt thresholds of 12, 5, and 5\GeV, and a triple tracker muon trigger with \pt thresholds of 12, 10, and 5\GeV. This last trigger requires ``tracker muons'' for which the tracker muon algorithm matches candidate muon segments from the tracker to segments from the innermost muon station~\cite{CMS:2018rym}.

The 2018 data set retains the last three triggers used in the 2017 data set but replaces the first trigger by one that is sensitive to displaced signatures. The so-called ``double standalone'' (SA) muon trigger requires at least two muons reconstructed using the muon detectors only~\cite{CMS:2018rym}, with a transverse momentum and pseudorapidity of $\pt > 23\GeV$ and $\abs{\eta} < 2$, respectively. This trigger was operated in 2018 and is used in the 2018 analysis. This double SA muon trigger does not rely on a primary vertex (PV) constraint for the track fit and is sensitive to both the prompt and displaced muons probed in this search. The PV is taken to be the vertex corresponding to the hard scattering location of the original pp collision, evaluated using tracking information alone~\cite{CMS-TDR-15-02}. There was no equivalent trigger in 2017; a study on 2017 data of the comparable double SA muon triggers available in 2018 either showed poor efficiencies for displaced muons in the pixel volume of the CMS experiment or negligible effective luminosities, thereby eliminating the possibility of investigating displaced track and vertex signatures in the 2017 analysis.

For both the 2017 and 2018 analyses, we require at least four offline reconstructed muons in each event. Among these muons, the 2017 analysis requires all muons to be reconstructed with the particle-flow (PF) algorithm~\cite{CMS:2017yfk}, which performs a global fit that combines information from each subdetector, while the 2018 analysis requires at least three muons to be reconstructed with the PF algorithm and at most one SA muon reconstructed using only the information from the muon system. Selecting one possible SA muon in the 2018 analysis mitigates the efficiency lost for displaced muon reconstruction in the tracker by roughly 30\%. Additional offline selection criteria are as follows: Each muon in the 2017 and 2018 analyses must have $\pt > 8\GeV$ and $\abs{\eta} < 2.4$. For the 2017 analysis, we require at least two ``high-\pt muons'', \ie, muons with $\pt > 13\GeV$. In the 2018 analysis, we require a higher threshold for these two ``high-\pt muons'', with $\pt > 24\GeV$. For both the 2017 and 2018 analyses, we also require that these ``high-\pt muons'' are located in the region $\abs{\eta} < 2$. 

For both the 2017 and 2018 analyses the following selections are made: 
Any two oppositely charged muons in the event are paired as long as their invariant mass is below 60\GeV. The decay vertices of each dimuon in each event are reconstructed using the Kalman filtering (KF) technique~\cite{CMS:2014pgm}, and only those vertices with a valid fitted vertex are retained. We then select the two dimuons closest in invariant mass. These two dimuons must not have any muons in common with each other. A dimuon that does not contain a high-\pt muon is labeled a low-\pt dimuon (denoted as \dimuonTwo), while the other dimuon, which includes at least one high-\pt muon, is called a high-\pt dimuon (denoted as \dimuonOne). In the scenario where both dimuons contain a high-\pt muon, the dimuons are randomly labeled to prevent bias in the kinematic distributions. All single muons not included in the two dimuons are called orphan muons. No requirements are applied to the orphan muons. We require that each of the two dimuons contain at least one muon that has at least one valid hit in any of the layers of the pixel detector. This requirement ensures the selected dimuons originate from the signal bosons that decay in the pixel detectors.

To further ensure that the signal muons in each dimuon decay from the same boson, we require a limit on the Kalman-fitted dimuon vertex probability, \Pmumu. For the 2017 analysis, this level is set to 15\%. As the dimuon transverse displacement (\Lxy) and the opening angle of the two muons \mbox{$\big(\DR = \sqrt{\smash[b]{(\Delta\eta)^2+(\Delta\phi)^2}}\big)$} increases, the fitted vertex probability decreases. As the 2018 analysis considers displaced signals, we account for this difference by defining a probability threshold that is a function of \Lxy, \DR, and the number of SA muons, \Nsa, and select valid fitted vertices based on the criterion $\Pmumu > P\big(\Lxy, \DR, \Nsa\big)$. 

The model independence of the results in this search is confirmed by verifying that the ratio of the full reconstruction efficiency \epsilonFull over the generator level acceptance \alphaGen is independent of all signal benchmarks. The acceptance \alphaGen is defined as the fraction of MC-generated signal events that pass the generator-level kinematic and geometric selection criteria. The parameter \epsilonFull is defined as the fraction of MC-generated events that pass the trigger and full event selection criteria. The valid-vertices criterion is particularly important, as this selection contributes to the overall uniformity of the distribution in \EpsilonOverAlpha. The ratio \EpsilonOverAlpha is found to be insensitive to variations in all signal benchmark parameters: over all of the MC samples, an average value of $0.403 \pm 0.001$ ($0.418 \pm 0.001$) for \EpsilonOverAlpha is obtained for the 2017 (2018) analysis, along with its statistical uncertainty.

Finally, we place a selection on the transverse displacement of $\Lxy < 16\cm$ and on the longitudinal displacement of $\Lz < 51.6\cm$. The upper bounds on \Lxy and \Lz correspond to the dimensions of the outermost layers of the CMS pixel system and define the volume in which the decay of a new boson can be probed by this analysis.

We also require that each dimuon is sufficiently isolated by having a total isolation sum of less than 2.3\GeV. This isolation sum, $\text{Iso}_{\dimuon} = \sum_{\text{tracks}}\pt(\text{track})$, is calculated as the scalar sum of the transverse momenta of all reconstructed tracks with $\pt > 0.5\GeV$ in the vicinity of the dimuon, \ie, within $\DR < 0.4$ and $\abs{\Ztrack - \Zdimuon} < 0.1\cm$. Here, \Ztrack is defined as the $z$ coordinate of the point of closest approach of the track to the primary vertex along the beam axis, while \Zdimuon is the $z$ position of the vertex associated with the dimuon propagated back to the beamline along the dimuon direction vector. Tracks included in the dimuon reconstruction are excluded from the isolation calculation. Requiring $\text{Iso}_{\dimuon} < 2.3\GeV$ effectively removes about 72\% of the quantum chromodynamic (QCD) background radiation. For a comprehensive and compact representation of all selection criteria, see Table~\ref{tab:selection}.

\begin{table}[!htp]
\centering
\topcaption{\label{tab:selection}The event selection requirements for the 2017 and 2018 analyses. In the signal muon selection row, the particle-flow loose muons refer to those muons that have tracks in both the tracker and the muon system, which is contrasted with the standalone (SA) muon selection, which only requires tracks in the muon system.}
\resizebox{0.9\textwidth}{!}{
\begin{tabular}{llrr}
\hline
\multirow{2}{*}{Selection}                                                       & \multirow{2}{*}{\begin{tabular}[c]{@{}l@{}}Additional\\ information\end{tabular}} & \multicolumn{2}{c}{Requirement}                                                                                                               \\
                                                                                 &                                                                                   & \multicolumn{1}{c}{2017}                                   & \multicolumn{1}{c}{2018}                                                         \\ \hline
\begin{tabular}[c]{@{}l@{}}Signal muon\\ candidates\end{tabular}                 & All 4 signal muons                                                                 & \begin{tabular}[c]{@{}r@{}}4 PF loose muons\end{tabular} & \begin{tabular}[c]{@{}r@{}} $\geq$3 PF loose muons\\ and $\leq$1 SA muon\end{tabular} \\
\multirow{2}{*}{\pt ($\abs{\eta}$)}                                          & All 4 signal muons                   &  $\pt > 8\GeV$  ($\abs{\eta} < 2.4$)                                                        & $\pt > 8\GeV$ ($\abs{\eta} < 2.4$)                                                                            \\                    
                                                                                 & 2 signal muons                   & $\pt > 13\GeV$ ($\abs{\eta} < 2.0$)                                                          & $\pt > 24\GeV$ ($\abs{\eta} < 2.0$)                                                                                \\
                                                                                
\begin{tabular}[c]{@{}l@{}}Invariant mass\end{tabular}                         & Each dimuon                             & $m_{\dimuon_{1,2}} < 60\GeV$                                                            & $m_{\dimuon_{1,2}} < 60\GeV$                                                                                  \\
\begin{tabular}[c]{@{}l@{}}Fitted dimuon\\ vertex probability\end{tabular} & Each dimuon                            & $P_{\dimuon_{1,2}} > 0.15$                                                           & $P_{\dimuon_{1,2}} > P(\Lxy,\DR,\Nsa)$                                                                                  \\
Dimuon isolation                                                                 & Each dimuon                            & \begin{tabular}[c]{@{}r@{}}$\text{Iso}_{\dimuon_{1,2}} < 2.3\GeV$ \\ ($\DR < 0.4$)\end{tabular}        & \begin{tabular}[c]{@{}r@{}}$\text{Iso}_{\dimuon_{1,2}} < 2.3\GeV$ \\ ($\DR < 0.4$)\end{tabular}                                                                                 \\
Fiducial volume                                                                  & Each dimuon                             & \multicolumn{1}{r}{\NA}                                       & \begin{tabular}[c]{@{}r@{}}$\Lxy < 16.0\cm$ \\ $L_{z} < 51.6\cm$ \end{tabular}                                                            \\ \hline
\end{tabular}}
\end{table}

\section{Signal shape modeling and signal region definition}\label{sec:signal_shape}
The target new boson is presumed to be weakly coupled to SM particles with a narrow width; therefore, the shape of the dimuon invariant mass distribution is fully determined by the detector resolution and final-state radiation (FSR) from the muons.

The signal region (SR), defined by the signal mass window in the two-dimensional (2D) plane of \MdimuonOne and \MdimuonTwo, is determined by fitting the prompt signal shape distributions of the MC signal samples with a double-sided Crystal Ball (CB) function (modified from~\cite{Gaiser:1985ix} to include a tail on each side of the Gaussian function). The fitting results for the displaced signal show no significant difference within the statistical fluctuations. This composite function consists of a Gaussian core and two power-law tails. The sigma parameter of the CB function describes the width of the Gaussian fit to the signal and, consequently, the signal resolution. To set the mass window size across the signal mass phase space, we extract the sigma parameter from the mass fit for each simulated mass point and plot it as a function of the invariant dimuon mass. To create a continuous signal mass window, we use the values obtained by linearly interpolating the plotted data points.

\begin{figure}[!htp]
\centering
\includegraphics[width=0.5\textwidth]{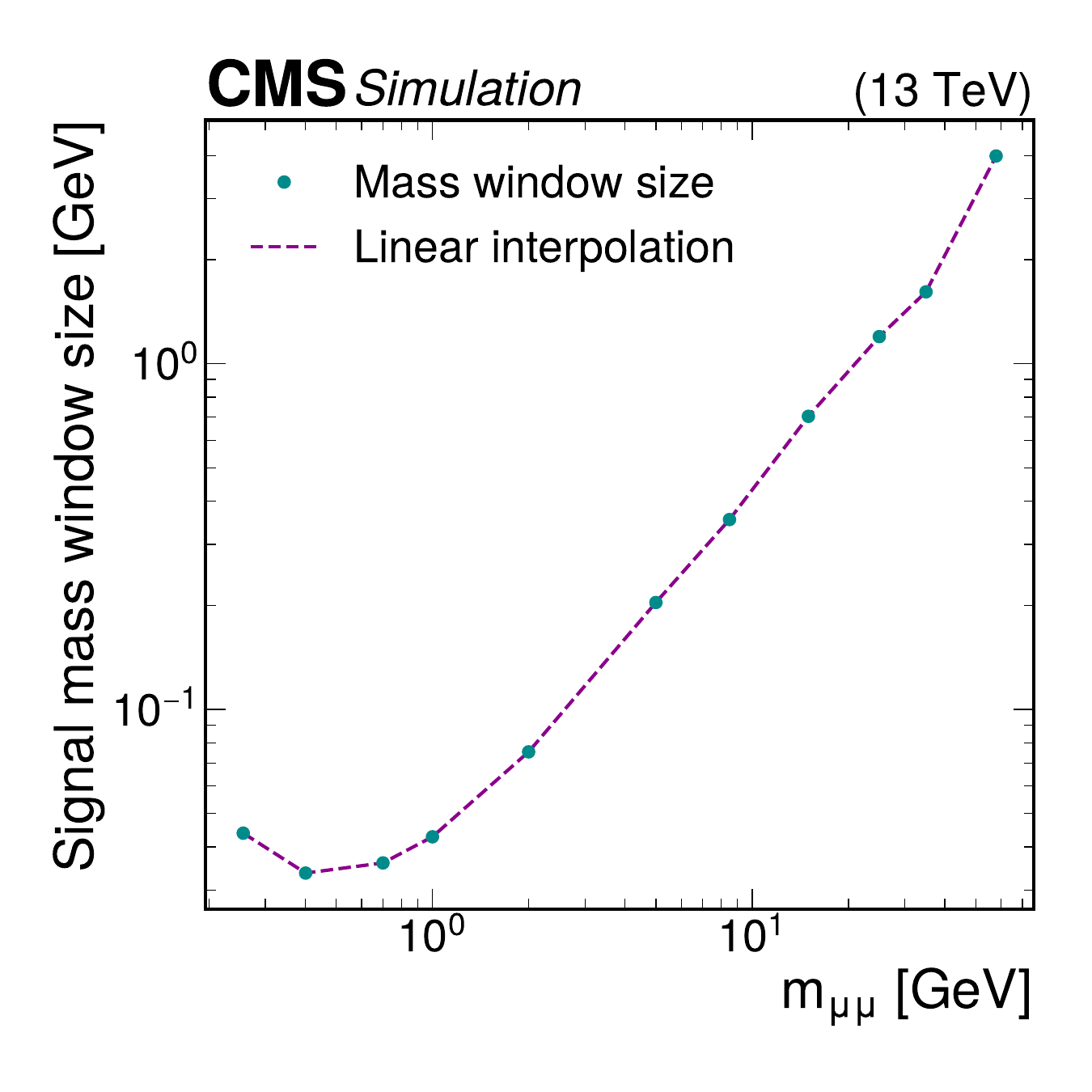}
\caption{\label{fig:masswindow}The mass window size as a function of invariant dimuon mass. It is derived from a Crystal Ball function fitting to MC signal events to contain 90\% of events. The wider mass window size around $m_{\PGm\PGm} \lesssim 0.4\GeV$ is due to the deteriorating mass resolution for near-collinear dimuons in the decays of low-mass bosons.}
\end{figure}

Because the two dimuons originate from identical bosons, the signal mass window selection is motivated by the requirement of consistent invariant masses of the dimuons.
This requirement is displayed in Equation~\eqref{eqn:consistent_masses} below,
\begin{equation}\label{eqn:consistent_masses}
\abs{\MdimuonOne - \MdimuonTwo} < W\big((\MdimuonOne + \MdimuonTwo)/2\big)
\end{equation}
where $W$ is a function of the dimuon invariant masses \MdimuonOne and \MdimuonTwo, and is based on the interpolation of the sigma parameters previously discussed. To determine the best signal significance, $S / \sqrt{S + B}$, where $S$ is the signal and $B$ is the expected background in the SR, thereby maximizing the discovery potential of this work, we examine the effect of four different signal efficiencies for prompt signals: 80\%, 85\%, 90\%, and 95\%.

\begin{figure}[ht!]
\centering
\includegraphics[width=0.46\textwidth]{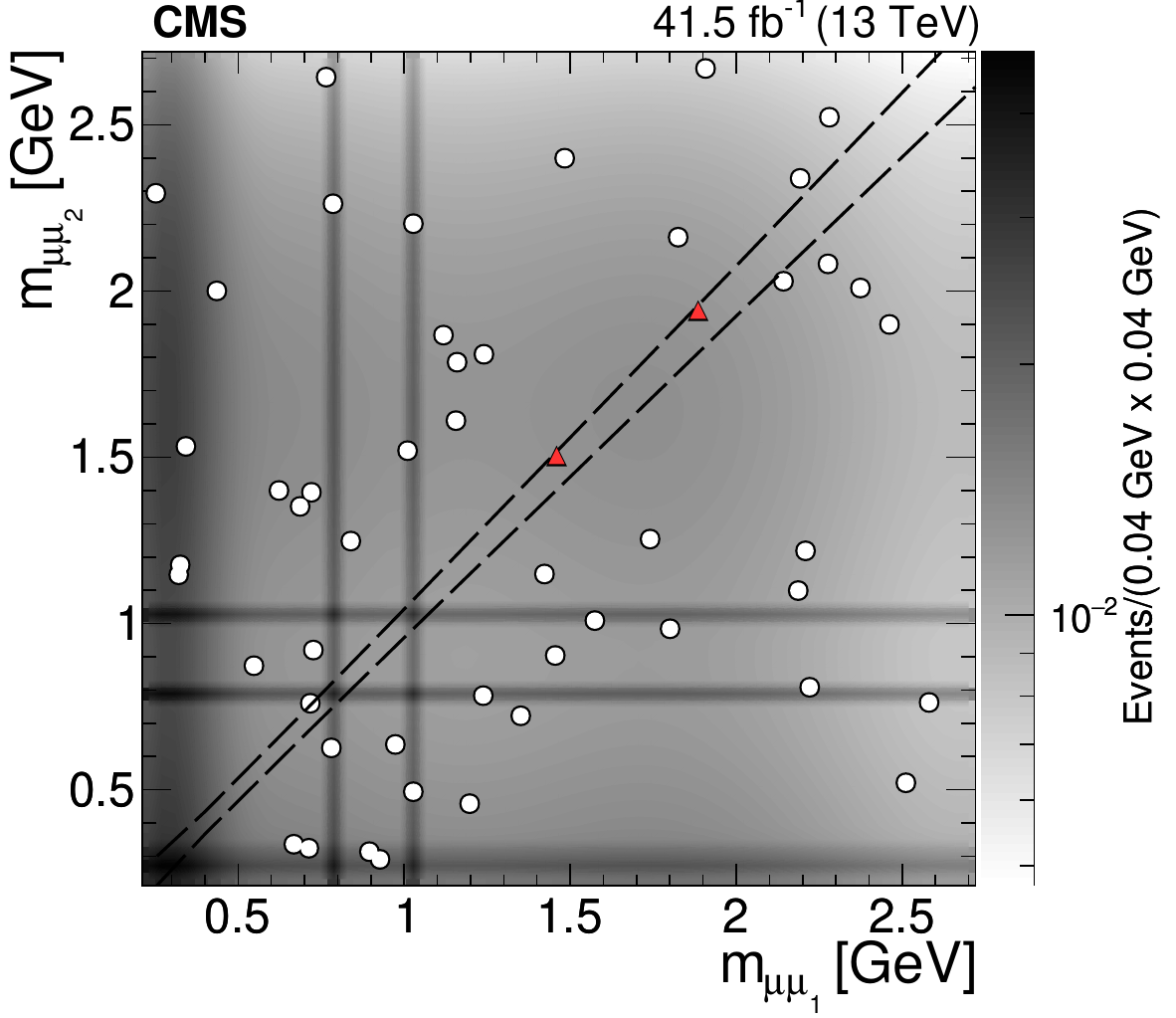}
\hfill
\includegraphics[width=0.46\textwidth]{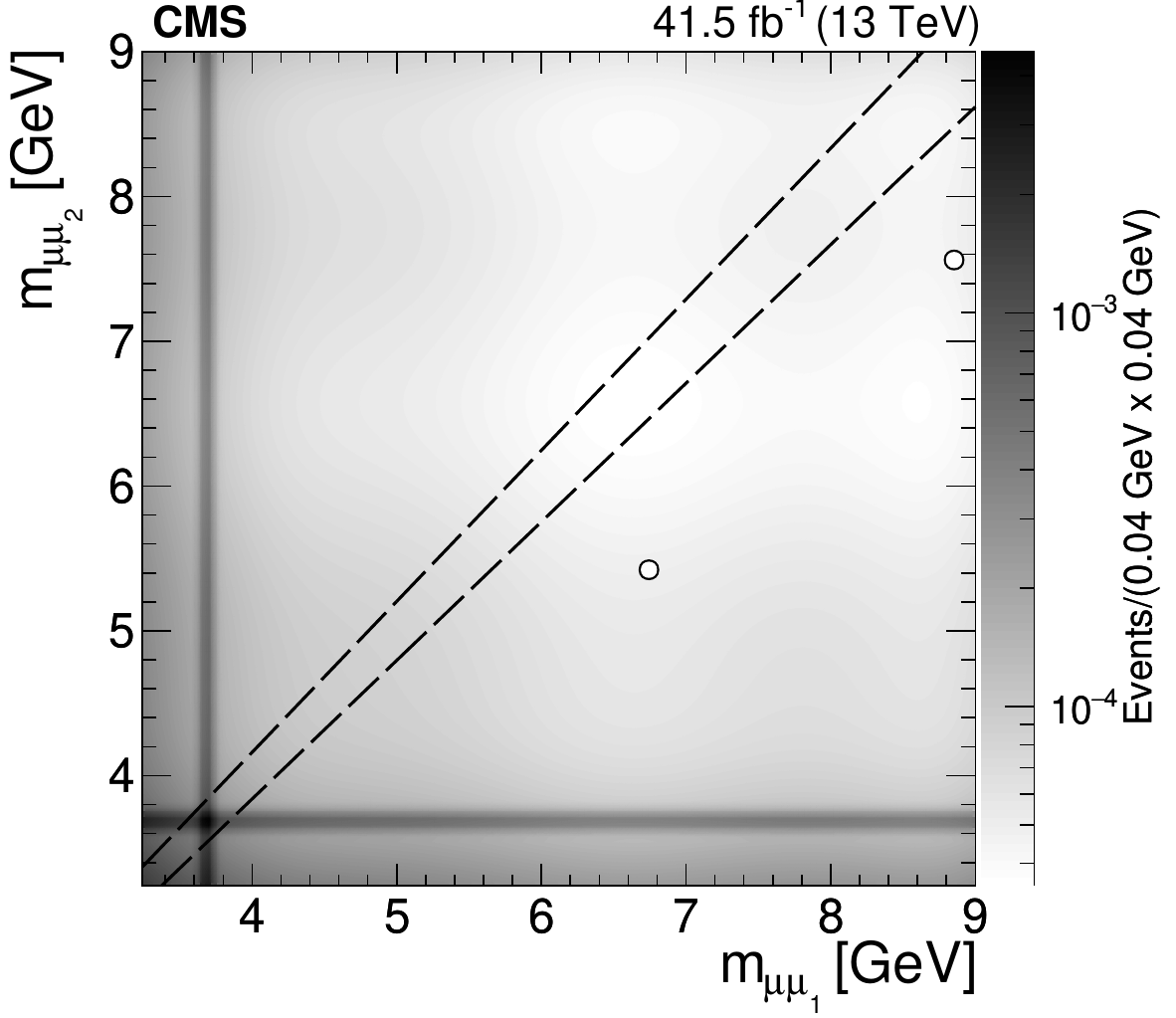}\\
\includegraphics[width=0.46\textwidth]{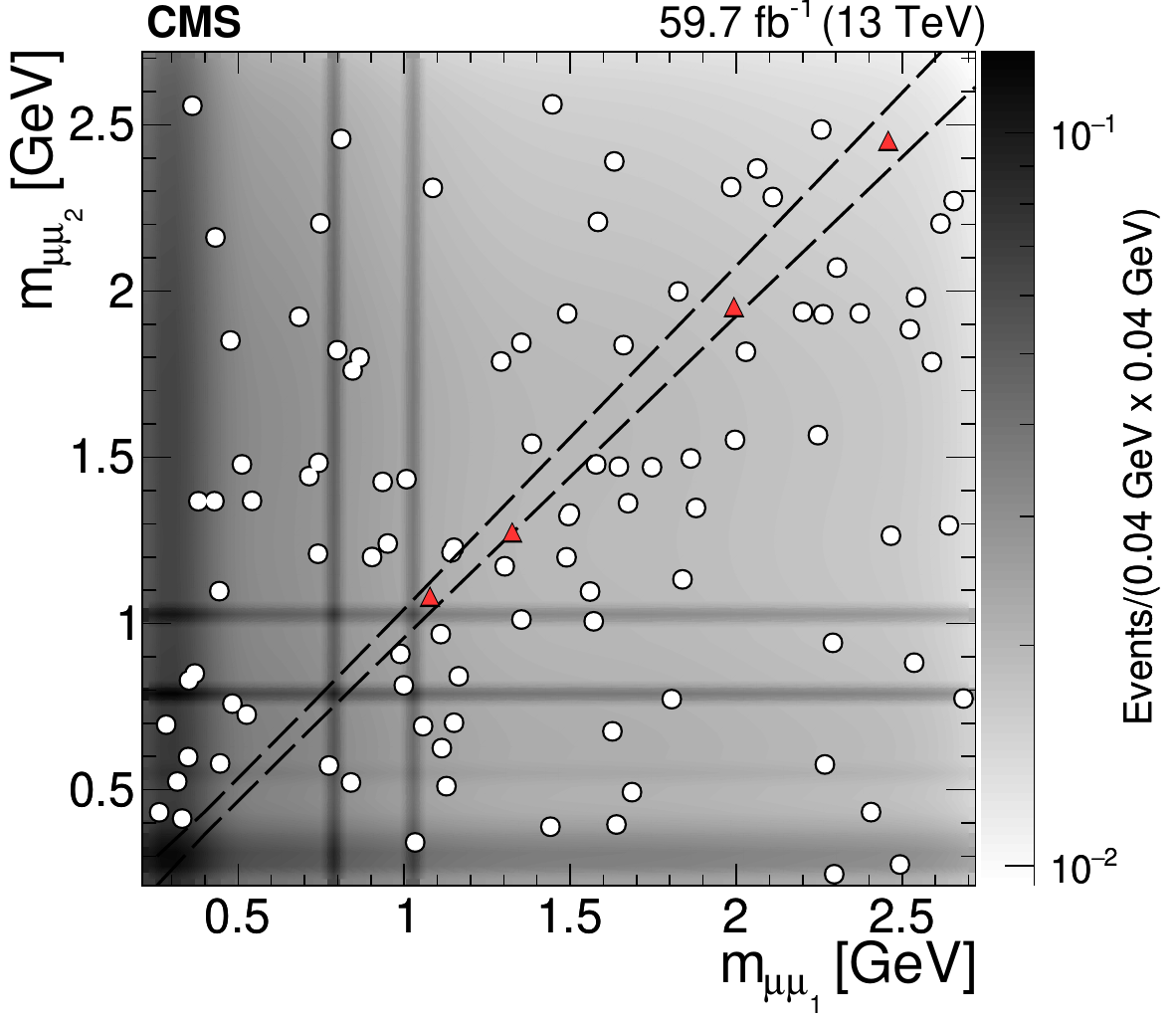}
\hfill
\includegraphics[width=0.46\textwidth]{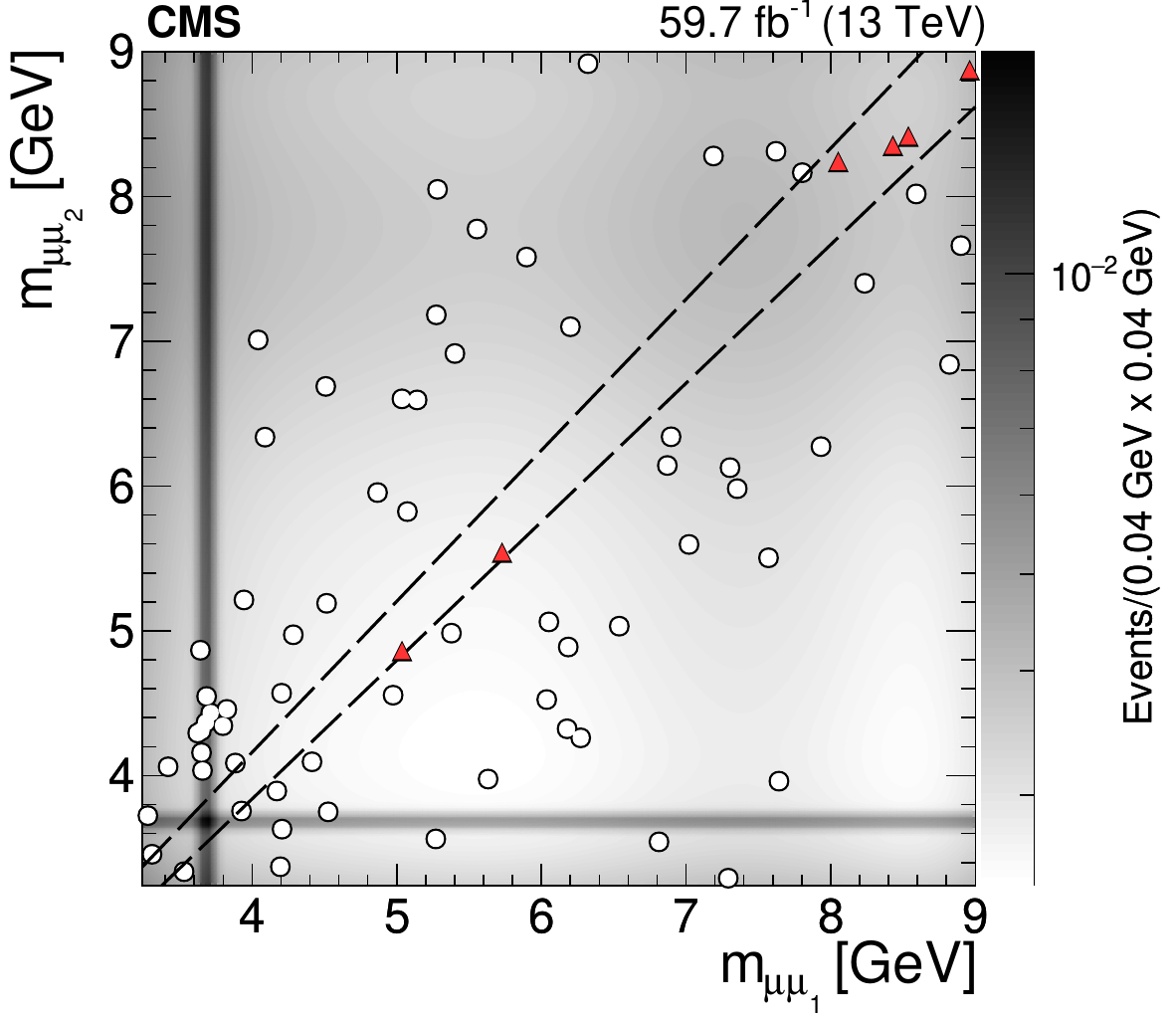}
\caption{Two-dimensional distribution of the invariant masses \MdimuonOne vs. \MdimuonTwo below (left) and above (right) the \JPsi resonance, for the 2017 (top row) and 2018 (bottom row) analyses. The greyscale heatmaps show the normalized QCD background templates, and the black vertical and horizontal bands correspond to the regions around the \Pgh, \Pgo,\Pgf, and \Pgy resonances. The white dots represent data events that pass all selection criteria but fall outside the SR $\MdimuonOne \simeq \MdimuonTwo$ (outlined by dashed lines), and the red triangles represent data events passing all selection criteria. As discussed in Section~\ref{subsec:bbbar}, the paucity of events in the CR for the 2017 analysis, particularly the region above \JPsi, is a result of the triggers selected for this data-taking period.}
\label{fig:UnblindingResultBelowUpsilon}
\end{figure}

\begin{figure}[ht!]
\centering
\includegraphics[width=0.48\linewidth]{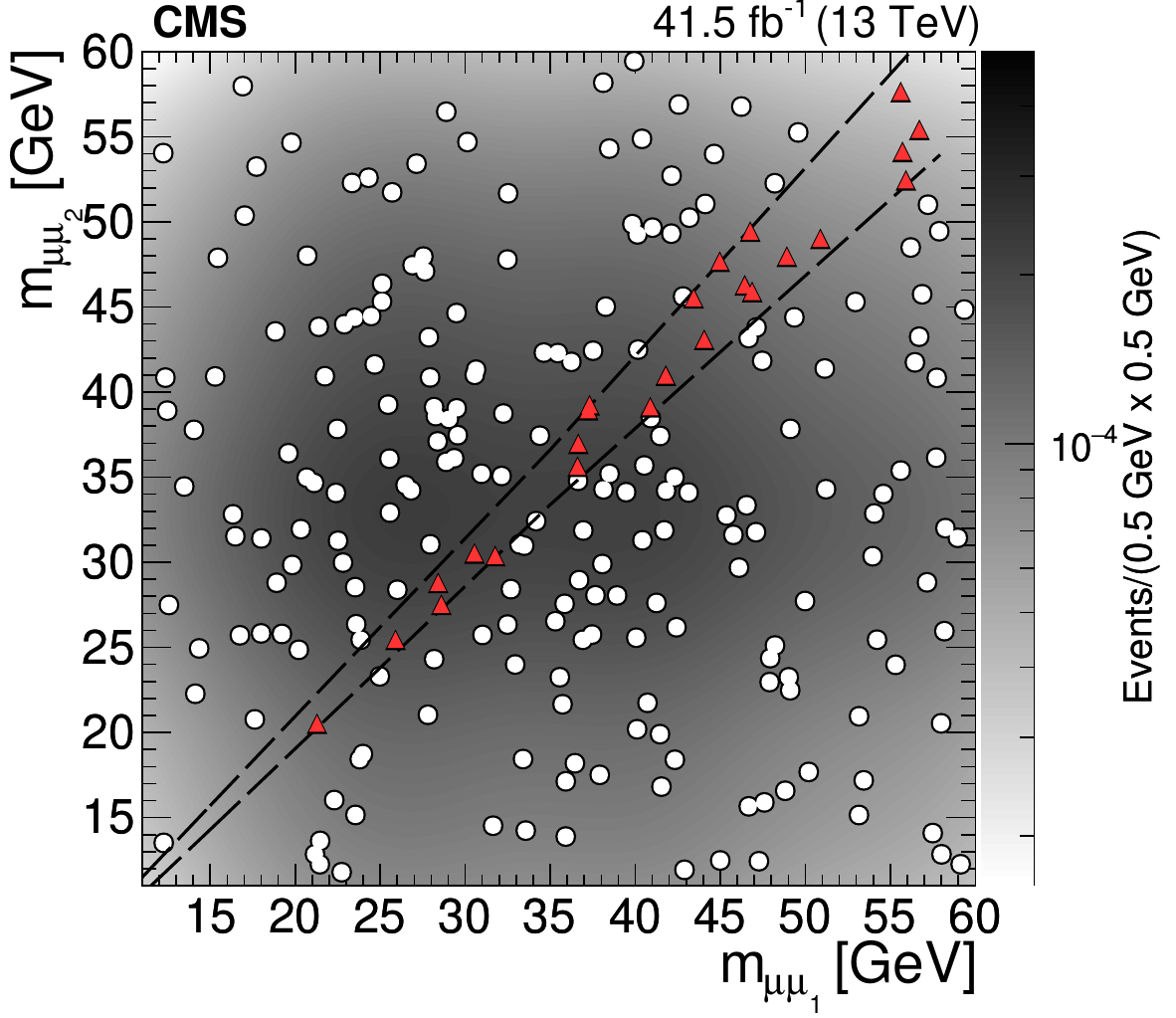}
\includegraphics[width=0.48\linewidth]{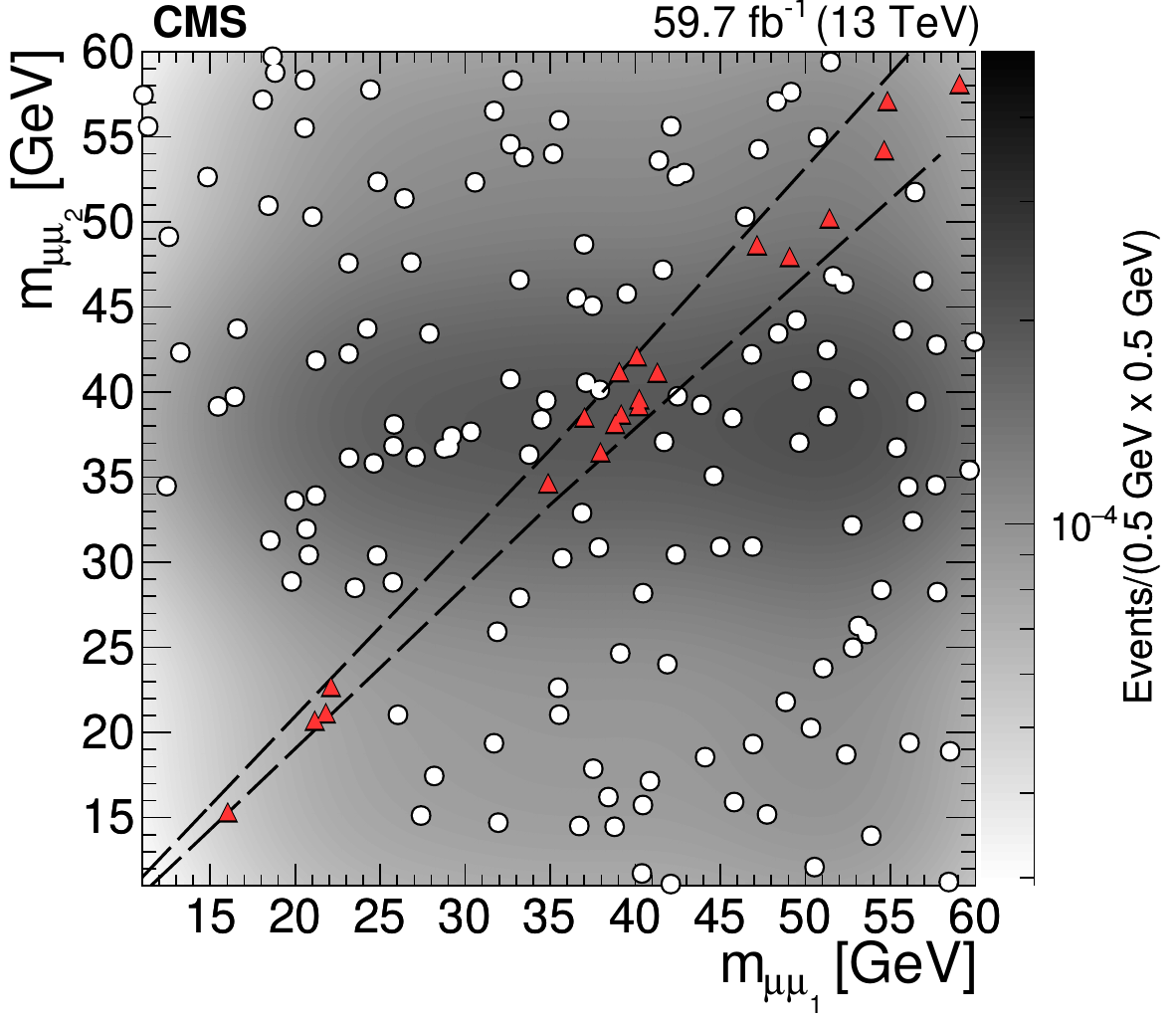}
\caption{Two-dimensional distribution of the invariant masses \MdimuonOne vs. \MdimuonTwo above the \PgU resonances for the 2017 analysis (left) and the 2018 analysis (right). The greyscale heatmaps represent the normalized 2D probability density function of the background data. White dots represent data events that pass all selection criteria but fall outside the SR $\MdimuonOne \simeq \MdimuonTwo$ (outlined by dashed lines), and the red triangles represent data events passing all selection criteria.}
\label{fig:UnblindingResultAboveUpsilon}
\end{figure}

For all signal efficiencies chosen, the signal significance decreases as the dimuon mass increases. However, over the entire mass range probed, $0.21 < m_{\PGm\PGm} < 60\GeV$, a signal mass window providing a 90\% efficiency on signal events yielded the best signal significance. Figure~\ref{fig:masswindow} displays the size of the signal mass window as a function of invariant dimuon mass (green dots) with the line derived from linear interpolation (dashed, purple line). Thus, given the combination of the mass window size and placement in the (\MdimuonOne, \MdimuonTwo) mass phase space, we achieve the optimal balance of signal significance and efficiency across the entire signal mass range, which we use to define the signal region (SR). This selection ultimately carves out the signal region in the 2D plane of \MdimuonOne and \MdimuonTwo, as illustrated by the dashed lines in Figs.~\ref{fig:UnblindingResultBelowUpsilon} and~\ref{fig:UnblindingResultAboveUpsilon}.

\section{Background estimation}\label{sec:bkg}
To estimate the background, we first divide the mass range into regions below and above the \PgU resonances and then split the range below the \PgU further at the location of the \JPsi resonance. We exclude the mass ranges that contain the \JPsi (2.72--3.24\GeV) and the \PgU (9--11\GeV) resonances. For the following discussion, we estimate the background in each region using fully data-driven methods based on data collected with the CMS detector.

\subsection{Background below the \texorpdfstring{\PgU}{Y} resonances}\label{subsec:bbbar}
The major background contribution below the \PgU resonances originates from QCD multijet processes. The QCD multijet background is dominated by events in which two \PQb quarks decay to $2\PGm + \PX$ or decay through low-mass meson resonances such as \PGo, \PGr, \PGf, and \Pgy. A two-dimensional template $T(\MdimuonOne, \MdimuonTwo)$ is constructed in the plane of the two dimuon invariant masses to estimate the QCD background. 

For both the 2017 and 2018 analyses the template is constructed over several steps. First, a control sample is selected from the data. Events are required to have only one dimuon, at least one orphan muon, and pass the event selection criteria as applied to signal candidates, including the vertex probability and isolation selection on the only dimuon, the pixel hit requirements, and all of the signal triggers. The primary HLT used in this study is a high-\pt, double muon trigger; consequently, the three muon selection requirement must contain two high-\pt muons. If an event includes more than one orphan muon, the muon with the highest \pt is selected. Depending on where the high-\pt muons are, these events are separated into two categories. The first category corresponds to events where both muons in the only dimuon are high-pT muons, the mass of this dimuon system is used to model the \MdimuonOne template. The second category contains events where the orphan muon is a high-\pt muon and the dimuon contains another high-\pt muon, the mass of the dimuon defines, in that case, the \MdimuonTwo variable. This procedure ensures that kinematic differences between signal events that have exactly two high-\pt dimuons or just one high-\pt dimuon are accounted for. For the 2018 analysis, the only dimuon in the event is required to have at most one SA muon, pass the selection on fitted vertex probability, and have at least one hit in the pixel system as explained in Section~\ref{sec:data_selection}.

Next, two one-dimensional (1D) dimuon mass templates, $T_{\mathrm{I}}(\MdimuonOne)$ and $T_{\mathrm{II}}(\MdimuonTwo)$, are obtained from the control sample. Each distribution is fitted with a function comprised of a Gaussian distribution for each light meson resonance and a set of sixth-degree Bernstein polynomials for the continuum background shape. The 2D template $T(\MdimuonOne, \MdimuonTwo)$ is obtained by taking the tensor product of the 1D templates, $T_{\mathrm{I}}(\MdimuonOne) \otimes T_{\mathrm{II}}(\MdimuonTwo)$.

Finally, the 2D template is normalized to unity in the dimuon-dimuon mass space for the region below and above the \JPsi resonance. The template is represented as a smooth function of the Cartesian product of \MdimuonOne and \MdimuonTwo, and is shown by the greyscale colormap in Fig.~\ref{fig:UnblindingResultBelowUpsilon}. Here, the SR defined in Section~\ref{sec:data_selection} is bounded by the dashed lines. The region of the mass space outside the SR represents the control region (CR) for the QCD background. The ratio between the integral of the template in the SR ($I_{\mathrm{SR}}$) and the CR ($I_{\mathrm{CR}}$), or $R = I_{\mathrm{SR}}/I_{\mathrm{CR}}$, is calculated for both the region below and above the \JPsi. For the 2017 analysis, the ratios in the region below and above the \JPsi are 0.046 and 0.096, respectively. Similarly, the ratios obtained for the 2018 analysis below and above the \JPsi are 0.044 and 0.093, respectively. The same figure also displays the events found in the data that pass all selection criteria except the consistent mass requirement, thus falling outside of the SR (represented as white dots). 

For the 2017 (2018) analysis, there are in total 49 and 2 (98 and 66) events in the CR below and above the \JPsi resonance, respectively. We then estimate the number of background events in the SR below and above the \JPsi using the ratio $R$. In 2017, these values are $2.26\pm 0.32\stat$ and $0.19\pm 0.14\stat$, below and above the \JPsi, respectively. In 2018, these values are $4.34\pm 0.44\stat$ and $6.16\pm 0.76\stat$, below and above the \JPsi, respectively. The statistical uncertainty is calculated as the product of the ratio $R$ with the square root of the number of CR events, or $R\sqrt{\smash[b]{N_{\mathrm{CR}}}}$. Since $R$ effectively normalizes this value, it is also referred to as the normalization uncertainty. These values are also listed in Table~\ref{tab:num_events} for convenience.

{\tolerance=1600
This method of estimating the QCD background is further validated by performing a \mbox{Baker--Cousins (BC) $\chi^2$} test~\cite{BAKER1984437} for the data in the CR and the constructed template. Both the 2D template $T(\MdimuonOne, \MdimuonTwo)$ and data events in the CR are projected into 1D histograms. The template integral is then scaled to the data entries. To calibrate the statistical test we randomly generated one million sets of pseudo-data based on the template. The $\chi^2$ statistic of the BC test is calculated for each pseudo-data set and the template to obtain a distribution of the $\chi^2$ values. The final calibrated p-value is obtained by calculating the $\chi^2$ between the actual data and the template and quoting the fraction above the observed $\chi^2$ statistic in the $\chi^2$ distribution. Because we are considering a total mass phase space trifurcated by the \JPsi and \PgU resonances in two separate analyses (2017 and 2018), we perform the BC test for each of these regions and for both analyses. All p-values for \MdimuonOne are contained in [0.043, 0.924], and all p-values for \MdimuonTwo are contained within [0.164, 0.889] for both the 2017 and 2018 analyses in the region below the \PgU resonances. Note that the lower range of p-values results from the paucity of background events in the 2017 data set. For a complete list of p-values, see Table~\ref{tab:pvalues}.
\par}

\subsection{Background above the \texorpdfstring{\PgU}{Y} resonances}\label{subsec:HighMassBKG}
Electroweak processes with two \PZ bosons, the \ttbar process, and the \mbox{Drell--Yan (DY)} process all contribute to the background above the \PgU resonances. Additionally, we consider the possibility of a radiated photon in the DY process that converts into a dimuon. Such an event could become a non-negligible background when a photon-converted muon is paired with an oppositely charged muon from the DY process. Most of these events, however, fail the requirement of consistent dimuon masses and reside in the CR. The limited number of such events in the data makes it difficult to model the shape of their distribution. To veto these events, the trailing mass $m^{\text{alt}}_{\text{trailing}}$ and trailing opening angle $\Delta R^{\text{alt}}_{\text{trailing}}$ from the alternative pairs are used. These variables are calculated as follows: once two dimuons are obtained from the normal muon pairing algorithm mentioned in Section~\ref{sec:data_selection}, we alternatively pair these muons by taking one muon from each dimuon and pairing it with the oppositely charged muon in the other dimuon; the smaller invariant mass of alternative pairs is $m^{\text{alt}}_{\text{trailing}}$ and the smaller opening angle of two muons in alternative pairs is $\Delta R^{\text{alt}}_{\text{trailing}}$. This alternative pairing restores dimuons that come from the DY process and the radiated photon. Compared to the dimuon from the DY process, the photon-converted dimuon usually has a smaller invariant mass and a smaller opening angle between the two muons. A final requirement of $m^{\text{alt}}_{\text{trailing}} > 3\GeV$, or $\DR^{\text{alt}}_{\text{trailing}} > 0.2$, is used to veto these background events.

The selection of events with at most one SA muon in two dimuons also causes extra background that is hard to model with MC samples. For example, in the \ttbar process, one SA muon from the \PQb quark decay can be paired with an oppositely charged muon from the \PW boson decay to form a dimuon. These dimuons are almost exclusively prompt. However, in the signal MC data, events that contain one SA muon often happen when the signal boson is displaced. To minimize the contamination of the actual data with these background events, we reject two-dimuon events containing one SA muon if both dimuons are prompt-like with $\Lxy < 0.1\cm$. Also, SA muons in background events are usually caused by underlying event activity or punch-through effects. Such SA muons have fewer reconstructed muon segments in the muon detector than those from signal events. We require the SA muon to have at least two valid muon segments, which further rejects the background caused by the SA muon.

Because the template method~\cite{CMS:2018jid} fails to describe the background above the \PgU resonances accurately, we use the kernel density estimate (KDE) method~\cite{Cranmer:2000du} to model the background in the region above the \PgU for both the 2017 and 2018 data sets. The KDE method is specifically chosen because it is a non-parametric method that operates on unbinned data, thereby eliminating the reliance on the input parameters typically used in the template method. The 2D probability density function (pdf) is built via the KDE method in two steps: First, the 1D pdf, using a Gaussian kernel with an adaptive bandwidth, is taken for both \MdimuonOne and \MdimuonTwo. The 2D pdf is then formed via the outer product of the pdfs for \MdimuonOne and \MdimuonTwo, \ie, $\hat{f}(\MdimuonOne,\ \MdimuonTwo)~=~f_{\mathrm{I}}(\MdimuonOne)~\otimes~f_{\mathrm{II}}(\MdimuonTwo)$, where $f_{\mathrm{I}, \mathrm{II}}(\Mdimuon)$ are the 1D pdfs. Figure~\ref{fig:UnblindingResultAboveUpsilon} displays the 2D pdfs in the form of greyscale colormaps for the 2017 analysis (left) and the 2018 analysis (right). The dashed lines bound the SR, \ie, SR$\big(\MdimuonOne \simeq \MdimuonTwo\big)$, and the white dots represent data events that fall outside this SR. The number of events in the CR above the \PgU resonances is 212 and 143 for the 2017 and 2018 data sets, respectively. In the mass region above the \PgU resonances, the ratio $R$ is 0.085 (0.097) for the 2017 (2018) analyses. We then estimate the number of expected background events in the SR. For 2017, we expect $18.10\pm 1.23\stat$ events, and for 2018, we expect $13.81\pm 1.16\stat$ events. The systematic uncertainties are explained in detail in Section~\ref{sec:syst}. For comparison across years and mass regions, these values and those from the low-mass regions are listed in Table~\ref{tab:num_events} and a summary of the background estimation and mitigation method across mass regions is listed in Table~\ref{tab:bkgest}.

\begin{table}[!htp]
\topcaption{\label{tab:num_events}The number of observed events in the CR and the expected and observed number of events in the SR in the three mass regions considered in this analysis.}
\resizebox{\textwidth}{!}{
\begin{tabular}{llrr}
\hline
\multirow{2}{*}{Region}                                                              & \multirow{2}{*}{Quantity} & \multicolumn{2}{c}{Year}                                                  \\
                                                                                     &                           & \multicolumn{1}{c}{2017}            & \multicolumn{1}{c}{2018}            \\ \hline
\multirow{2}{*}{Below \JPsi}                                                         & Obs. events in CR         & 49                                  & 98                                  \\
                                                                                     & Exp. events in SR         & $2.26\pm0.32\stat\pm 0.14\syst$                & $4.34 \pm0.44\stat\pm 0.18\syst$                 \\
                                                                                     & Obs. events in SR         & 2                                   & 4 \\
\hline
	\multirow{2}{*}{\begin{tabular}[c]{@{}l@{}}Above \JPsi\\ \& Below \PgU\end{tabular}}   & Obs. events in CR         & 2                                   & 66                                  \\
                                                                                     & Exp. events in SR         & $0.19\pm0.14\stat\pm 0.01\syst$                 & $6.16\pm 0.76 \stat\pm 0.09\syst$                 \\
                                                                                     & Obs. events in SR         & 0                                   & 6 \\
\hline                                                                                   
\multirow{2}{*}{Above \PgU}                                                          & Obs. events in CR         & 212                                 & 143                                 \\
                                                                                     & Exp. events in SR         & $18.10\pm1.23 \stat\pm 4.49 \syst$ & $13.81\pm 1.16\stat\pm5.39\syst$ \\ 
                                                                                     & Obs. events in SR         & 24                                  & 20 \\
                                                                                     \hline
\end{tabular}}
\end{table}

\begin{table}[!htp]
\centering
\caption{\label{tab:bkgest} The background estimation and mitigation method are divided into two mass regions below and above the \PgU resonance. The region below the \PgU uses the template method to estimate the background PDFs with signal-like three-muon events, which include only one dimuon and one orphan muon. The region above the \PgU uses the KDE method with events in the control region and involves further event selection for background mitigation.}
\begin{tabular}{l l l}
\hline
Region       &  Method & Selection \\
\hline
\multirow{1}{*}{Below \PgU} & \multirow{1}{*}{Template} & Signal-like three-muon events\\
\hline
\multirow{2}{*}{Above \PgU} & KDE & CR events \\
                           & Background mitigation & Veto $m^{\mathrm{alt}}_{\mathrm{trailing}} > 3\GeV$ or $\Delta R^{\mathrm{alt}}_{\mathrm{trailing}} > 0.2$ events\\
\hline
\end{tabular}
\end{table}

To assess the goodness-of-fit of the pdfs produced via the KDE method, we perform the BC $\chi^2$ test. Just like the template method, the 2D pdfs are separated into the 1D pdfs individually representing \MdimuonOne and \MdimuonTwo. The BC test is then performed on each 1D pdf in the same manner as described in Section~\ref{subsec:bbbar}, the results of which yield p-values in the range of $[0.513,\ 0.974]$ for both invariant dimuon masses and both analyses. For completeness, we tabulate all p-values for both \MdimuonOne and \MdimuonTwo in all three mass regions for both the 2017 and 2018 analyses in Table~\ref{tab:pvalues}.

\begin{table}[!htp]
\centering
\topcaption{\label{tab:pvalues}Results of the Baker--Cousins $\chi^2$ test for goodness-of-fit between the modeled background and the actual background events in the control regions. For background below the \PgU resonances, the 2D background templates are projected into one dimension, and the test is conducted, comparing background data and template values, yielding two p-values per region per data set. A similar procedure is followed for background above the \PgU resonances: the 2D pdfs are projected into one dimension and the BC test is conducted, yielding two p-values for each dimuon invariant mass for both 2017 and 2018. Note that the low p-values for the mass region above $\JPsi$ and below \PgU for the 2017 data set result from the low number of background events in the data.}
\begin{tabular}{llrrr}
\hline
\multirow{3}{*}{Region}                  & \multicolumn{4}{c}{p-value}                                                                                                                                                   \\
                                         & \multicolumn{2}{c}{2017}                                                    & \multicolumn{2}{c}{2018}                                                                        \\
                                         & \MdimuonOne & \multicolumn{1}{l}{\MdimuonTwo} & \multicolumn{1}{l}{\MdimuonOne} & \multicolumn{1}{l}{\MdimuonTwo} \\ \hline
Below $\JPsi$                            & \multicolumn{1}{r}{0.333}  & 0.889                                          & 0.186                                          & 0.164                                          \\
Above $\JPsi$ \& below \PgU & \multicolumn{1}{r}{0.043}  & 0.172                                          & 0.924                                          & 0.375                                          \\
Above \PgU                & \multicolumn{1}{r}{0.974}  & 0.695                                          & 0.768                                          & 0.513                                          \\ \hline
\end{tabular}
\end{table}

\section{Systematic uncertainties}\label{sec:syst}
The leading source of experimental uncertainty in the signal prediction comes from the measurement of the integrated luminosity recorded by the CMS detector: during the 2017 (2018) data-taking period, this uncertainty was 2.3\%~\cite{CMS:2018elu} (2.5\%~\cite{CMS:2019jhq}). Other experimental uncertainties include the uncertainty on the final signal efficiency at high PU conditions (2.3\% in 2017 and 1.8\% in 2018), the selection of consistent masses of the two dimuons (0.24\% for both 2017 and 2018), the uncertainty in the signal trigger scale factor that corrects for different trigger efficiencies in the data and MC simulation (0.9\% in 2017 and 0.6\% in 2018), and the uncertainty in the modeling of the PU distribution (0.1\% in 2017 and 0.05\% in 2018). The uncertainty on the reconstruction of displaced tracks and vertices contributes to the total systematic uncertainty of the 2018 analysis only, and is quoted at 1.0\%, with 0.5\% contributing from each displaced dimuon vertex fit~\cite{CMS:2022fut}. The experimental uncertainties due to muon identification, dimuon isolation, and reconstruction of close muons in the tracker and the muon system can be found in Ref.~\cite{CMS:2018jid}, and are reproduced in Table~\ref{tab:uncertainties}. The theoretical uncertainties include the uncertainty in the PDFs, knowledge of the strong coupling constant \alpS, the renormalization and factorization scales, and the production cross section of the Higgs boson and its branching fractions---all of these uncertainties can be found in Ref.~\cite{CMS:2018jid} as well, and are also reproduced in Table~\ref{tab:uncertainties}.

We also consider systematic uncertainties for the estimated background. The dimuon isolation threshold described in Section~\ref{sec:data_selection} is varied by 5\% around the threshold to achieve a total variation of 10\%; the final number of expected background events in the SR is recomputed. The changes in the expected background events are quoted as a systematic uncertainty for the region below \JPsi (6.6\% in 2017 and 4.1\% in 2018), between \JPsi and \PgU (5.9\% in 2017 and 1.5\% in 2018), and above \PgU (1.2\% in 2017 and 2.3\% in 2018).

For the region above the \PgU resonances, we determine a systematic uncertainty on the pdf background shape by examining the number of estimated events for two scenarios in which the number of CR events is reduced by half. This is accomplished by applying selections to the events in the CR, which reduce their number by 50\% and comparing the estimated number of events in the SR resulting from these selections. For all selections placed, we first keep the SR constant and then impose two different linear inequalities that eliminate roughly half of the events in the CR, which are given by $0 \leq\zeta\MdimuonOne - \theta\MdimuonTwo$ and $0 \geq \theta\MdimuonOne - \zeta\MdimuonTwo$, where $\zeta$ and $\theta$ are arbitrary parameters chosen such that the desired 50\% reduction of CR events is achieved.

 We use these inequalities to exclude data to determine how the number of expected events in the SR changes, which provides a measure of the uncertainty on the background pdf shape. To produce this result, we consider two scenarios for both the 2017 and 2018 analyses: for the first scenario, we set the parameters $\zeta = 1.6$ and $\theta = 1$, and determine the number of expected events in the SR. For the second scenario, we swap the values of $\zeta$ and $\theta$ and again determine the number of events in the SR. The ratio of the SR and CR background yields, \ie, $I_{\mathrm{SR}}/I_{\mathrm{CR}}$, are computed and the expected background yields are compared to the nominal case where no selection is applied. The uncertainty on the background pdf shape is then taken as the largest percent difference, resulting in a 23.6\% and 36.7\% uncertainty for 2017 and 2018, respectively. Note that we do not include uncertainty on the background shape below the $\PgU$ resonances because this region was estimated using the template method. Table~\ref{tab:uncertainties} presents all uncertainties used in this analysis.

\begin{table}[ht!]
\centering
\topcaption{The systematic uncertainties for the 2017 and 2018 analyses, including the experimental uncertainties on the signal and background, as well as the theoretical uncertainties. The experimental uncertainties on the muon identification (ID), the dimuon isolation, and the reconstruction of close muons in both the tracker and the muon system have been reproduced from~\cite{CMS:2018jid}. Additionally, all theoretical uncertainties have been reproduced from~\cite{CMS:2018jid}.}
\label{tab:uncertainties}
\begin{tabular}{lrr}
\hline
Source of uncertainty                                      & \multicolumn{2}{c}{Value}                           \\
                                                           & \multicolumn{1}{c}{2017} & \multicolumn{1}{c}{2018} \\ \hline
\multicolumn{3}{l}{Experimental signal uncertainties}                                                            \\ \hline
Integrated luminosity                                      & 2.3\%                    & 2.5\%                    \\
Muon ID                                                    & $4\times0.6\%$           & $4\times0.6\%$           \\
Dimuon isolation                                           & $2\times0.1\%$           & $2\times0.1\%$           \\
Reco. of close muons in tracker (signal mass $< 9\GeV$)   & $2\times1.2\%$           & $2\times1.2\%$          \\
Reco. of close muons in muon system (signal mass $< 9\GeV$)& $2\times1.3\%$           & $2\times1.3\%$          \\
Muon HLT                                                   & 0.9\%                    & 0.6\%                    \\
Reconstruction of displaced track/vertex                   & \NA                      & $2\times0.5\%$           \\
PU distribution                                            & 0.1\%                    & 0.05\%                   \\
PU effect on signal efficiency                                                 & 2.3\%                    & 1.8\%                    \\
Dimuon mass consistency                                    & 0.24\%                   & 0.24\%                   \\ \hline
\multicolumn{3}{l}{Experimental background uncertainties below  $\JPsi$ (0.21--2.72\GeV)}                       \\ \hline
Normalization                                              & 14.2\%                   & 10.1\%                   \\
Systematic                                                 & 6.6\%                    & 4.1\%                    \\ \hline
\multicolumn{3}{l}{Experimental background uncertainties above $\JPsi$ and below $\PgU$ (3.24--9\GeV)}          \\ \hline
Normalization                                              & 73.5\%                   & 12.3\%                   \\
Systematic                                                 & 5.9\%                    & 1.5\%                    \\ \hline
\multicolumn{3}{l}{Experimental background uncertainties above $\PgU$ (11--60 \GeV)}                             \\ \hline
Normalization                                              & 6.9\%                    & 8.4\%                   \\
Systematic                                                 & 1.2\%                    & 2.3\%                    \\
Shape                                                      & 23.6\%                   & 36.7\%                   \\ \hline
\multicolumn{3}{l}{Theoretical signal uncertainties}                                                             \\ \hline
PDF + $\alpS$ + QCD scales                            & 8\%                      & 8\%                      \\
Higgs cross-section and BR\textsuperscript{a}              & 3.8\%                    & 3.8\%                    \\
NNLO Higgs \pt re-weighting\textsuperscript{a}              & 2\%                      & 2\%                      \\ \hline
\end{tabular}\\[0.4em]
\hspace*{-5.5cm}\textsuperscript{a}Uncertainty is not used in the vector portal model.
\end{table}

{\tolerance=1600
During the 2017 and 2018 data-taking periods, an issue affecting the L1 trigger was \mbox{detected~\cite{CMS:2021yvr, CMS:2020cmk}}, affecting muon candidates' bunch crossing assignment. Here, the limited time resolution of the muon detectors could erroneously register a hit and fire the L1 trigger. To assess the impact of this L1 trigger issue, we re-weight the signal MC samples by the muon \pt and $\eta$ and obtained a corrected value of \epsilonFull. From this re-weighting, we conclude that this L1 trigger issue has a negligible effect on \epsilonFull, causing an impact of less than 1\%.
\par}

\section{Results}\label{sec:results}
After applying all selection criteria to the data sample, 26 (30) events are found in the full mass range in SR for the 2017 (2018) analysis. Their distribution in the \MdimuonOne and \MdimuonTwo phase space is shown in Figs.~\ref{fig:UnblindingResultBelowUpsilon} and \ref{fig:UnblindingResultAboveUpsilon}. This result is consistent with the sum of all background estimates described in Section~\ref{sec:bkg}. Model-independent observed upper limits at a 95\% confidence level (\CL) are set on the product of the production cross section of pairs of new bosons, the square of the branching fraction to dimuons, and the acceptance. Limits are set using the \CLs method~\cite{Read:2002hq, Junk:1999kv}. The test statistic used is based on the logarithm of the likelihood ratio~\cite{CMS:2014fzn}. The systematic uncertainties and their correlations have been accounted for by profiling the likelihood with respect to the nuisance parameters for each value of the signal strength. The following results have been determined using the CMS statistical analysis tool \textsc{Combine}~\cite{CAT-23-001}.

\begin{figure}[h!]
\centering
\includegraphics[width=0.47\linewidth]{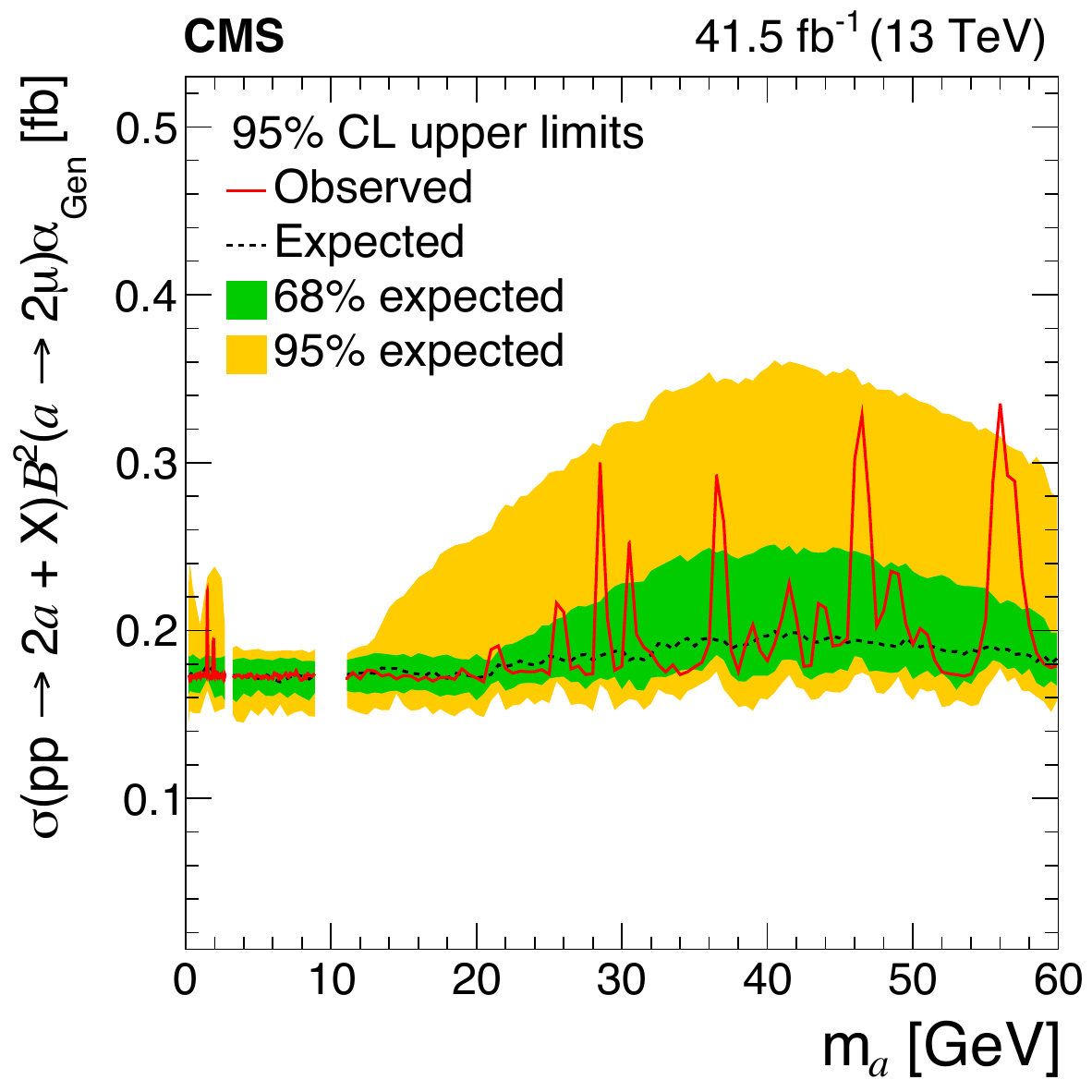}
\includegraphics[width=0.47\linewidth]{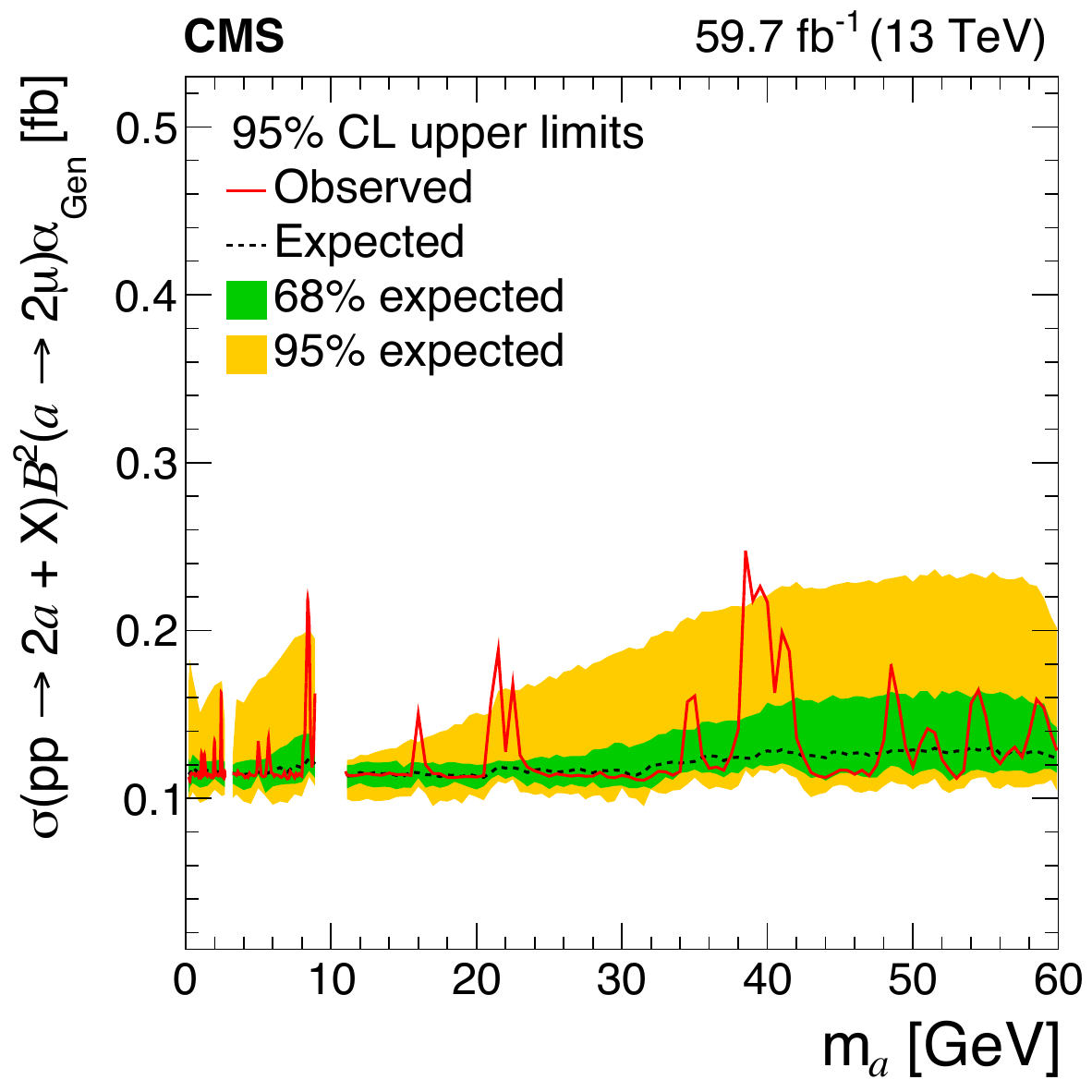}\\

\includegraphics[width=0.47\linewidth]{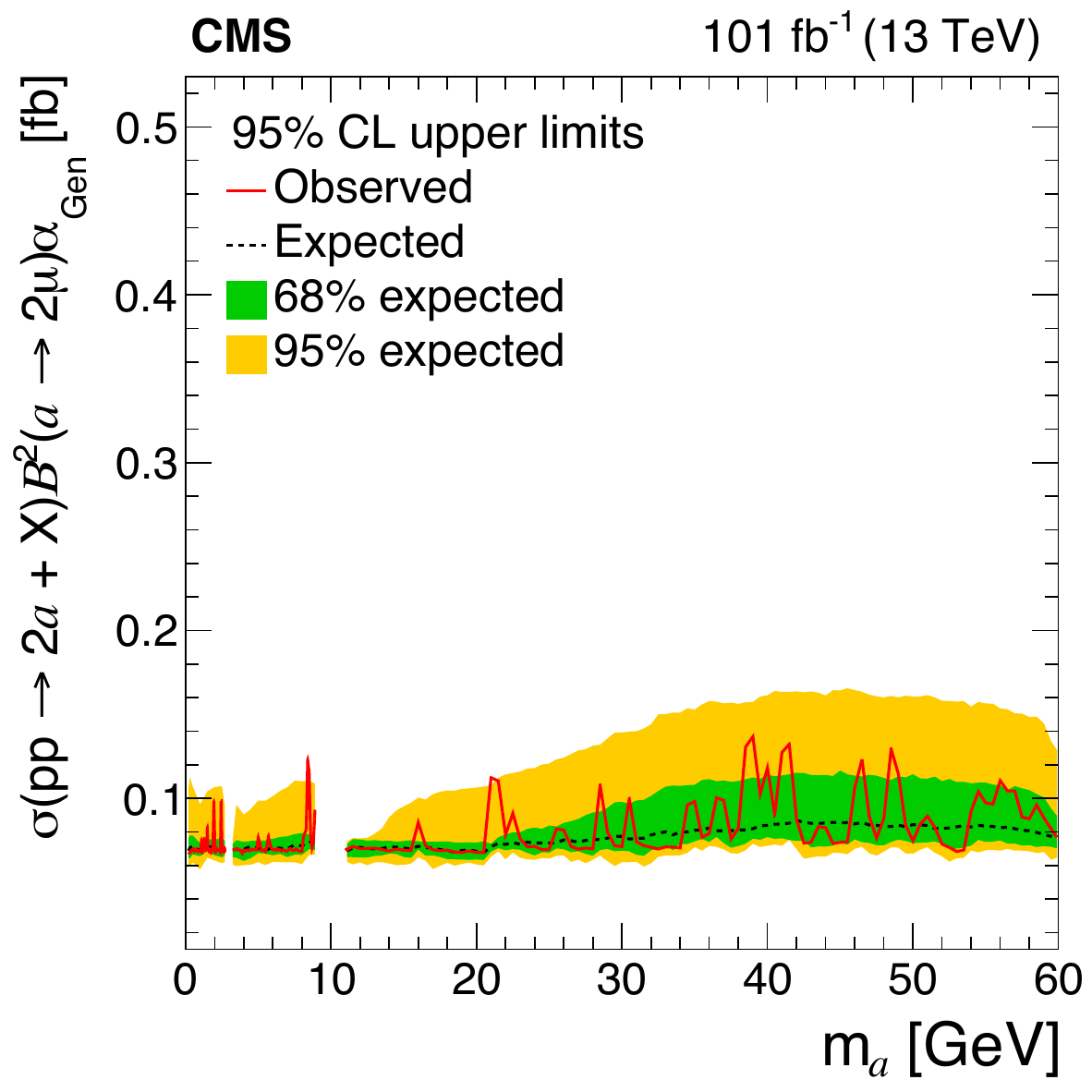}
\includegraphics[width=0.47\linewidth]{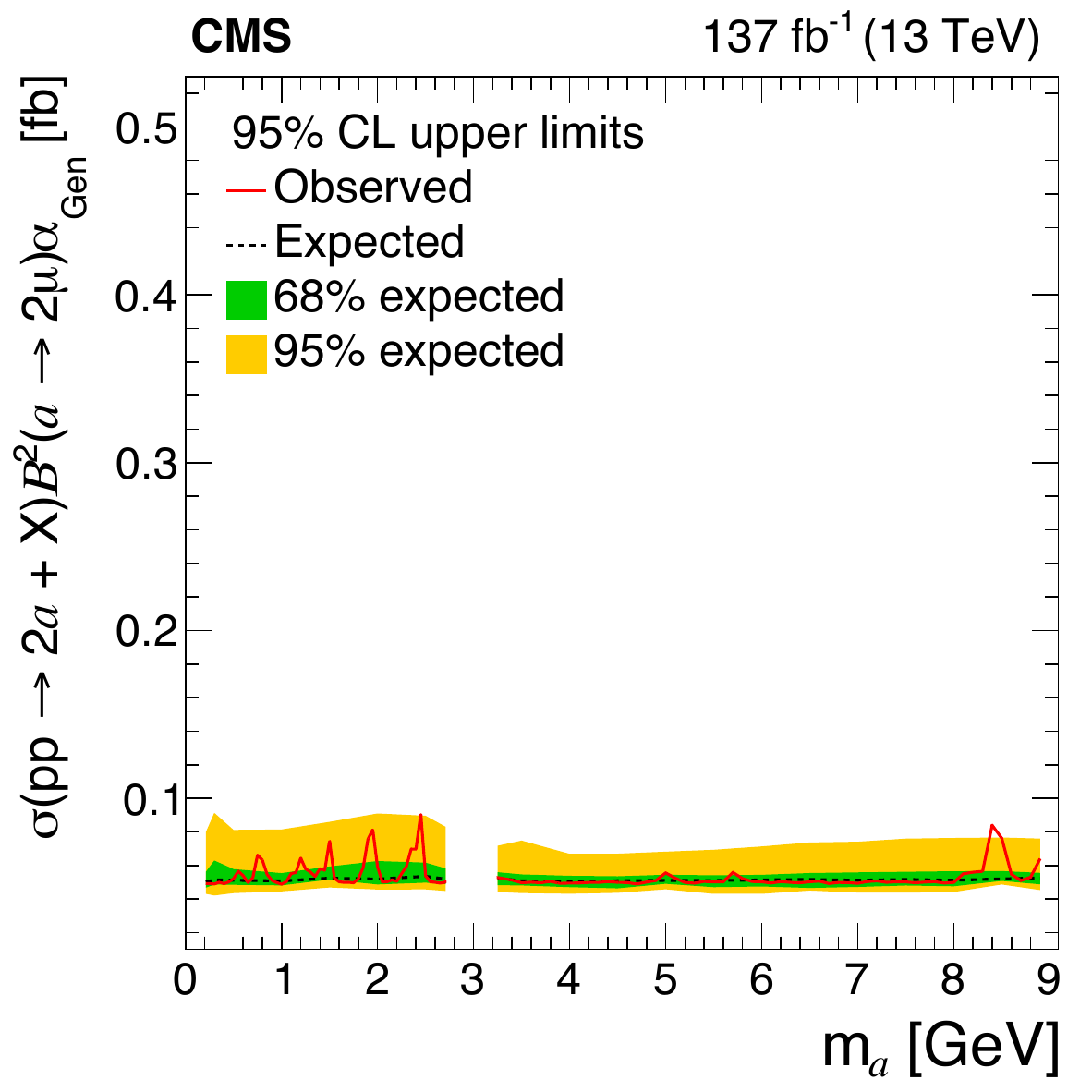}
\caption{The model-independent 95\% \CL expected and observed upper limits set on ${\sigma(\PP \to 2\Pa + \PX)\mathcal{B}^2(\Pa\to 2\PGm)\alphaGen}$ over the range $0.21 < \MPa < 60\GeV$ for the 2017 and 2018 analyses (top left and top right, respectively), over the range $0.21 < \MPa < 60\GeV$ for the combined 2017 and 2018 analyses (bottom left), and over the range $0.21 < \MPa < 9\GeV$ for the combined 2016, 2017, and 2018 analyses (bottom right). Mass ranges that overlap with \JPsi and \PgU resonances are excluded from the search.}
\label{fig:ModelIndependentLimit}
\end{figure}

{\tolerance=1600
The limit is shown as a function of \MPa in Fig.~\ref{fig:ModelIndependentLimit} over the range $0.21 < \MPa < 60\GeV$ for the 2017 and 2018 analyses (top left and top right, respectively), for the combined 2017 and 2018 data set over the range $0.21 < \MPa < 60\GeV$ (bottom left), and for the combined 2016, 2017, and 2018 data set over the range $0.21 < \MPa < 9\GeV$ (bottom right), where the analysis presented in Ref.~\cite{CMS:2018jid} provides the input data for the 2016 data set. Overall, the limit for the 2018 data set, the combination of the 2017 and 2018 data sets, and the combination of the 2016, 2017, and 2018 analyses varies between 0.049 and 0.247\unit{fb} for the model-independent interpretation. We now interpret this result in the context of the signal benchmark models.
\par}

\begin{figure}[h!]
\centering
\includegraphics[width=0.49\textwidth]{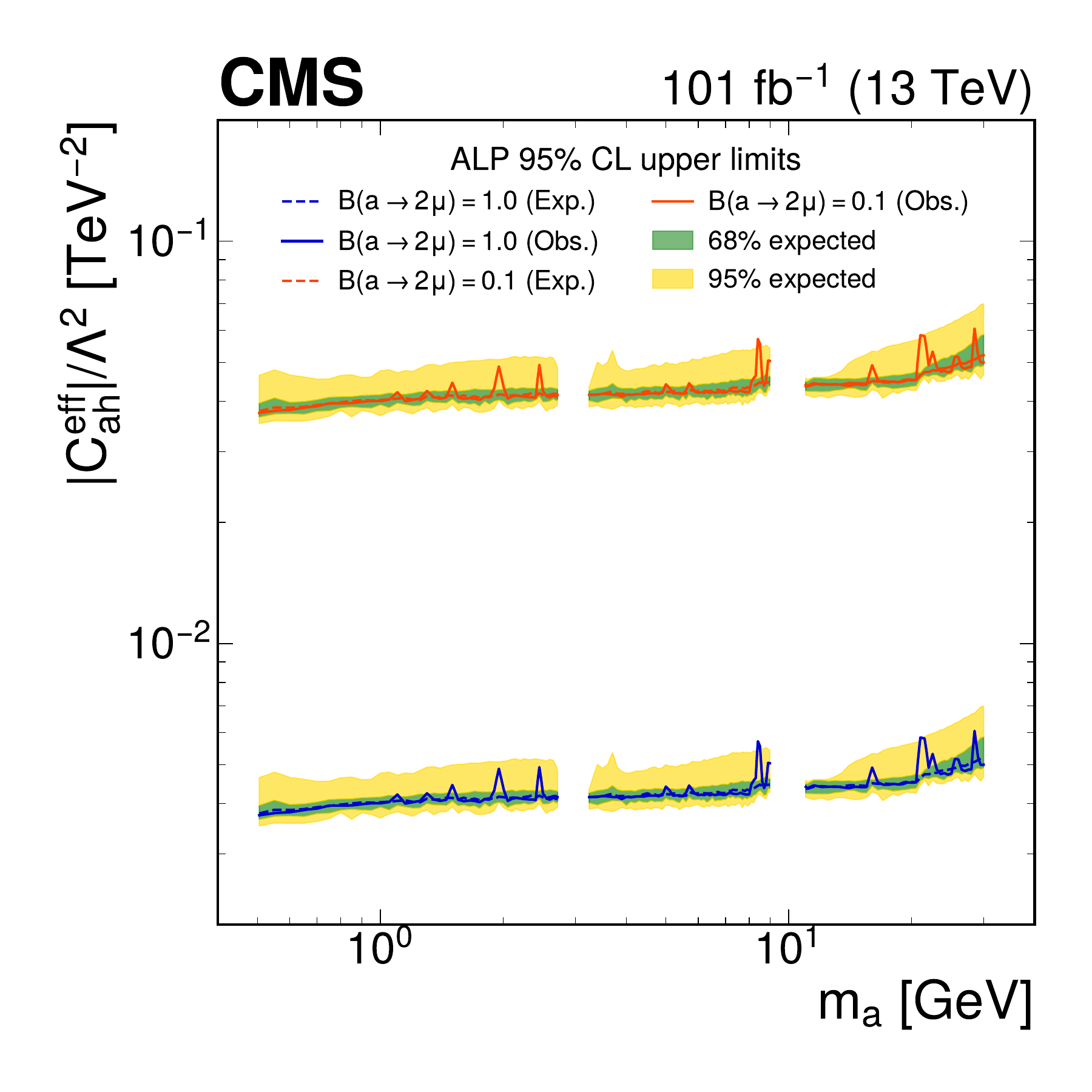}
\hfill
\includegraphics[width=0.49\linewidth]{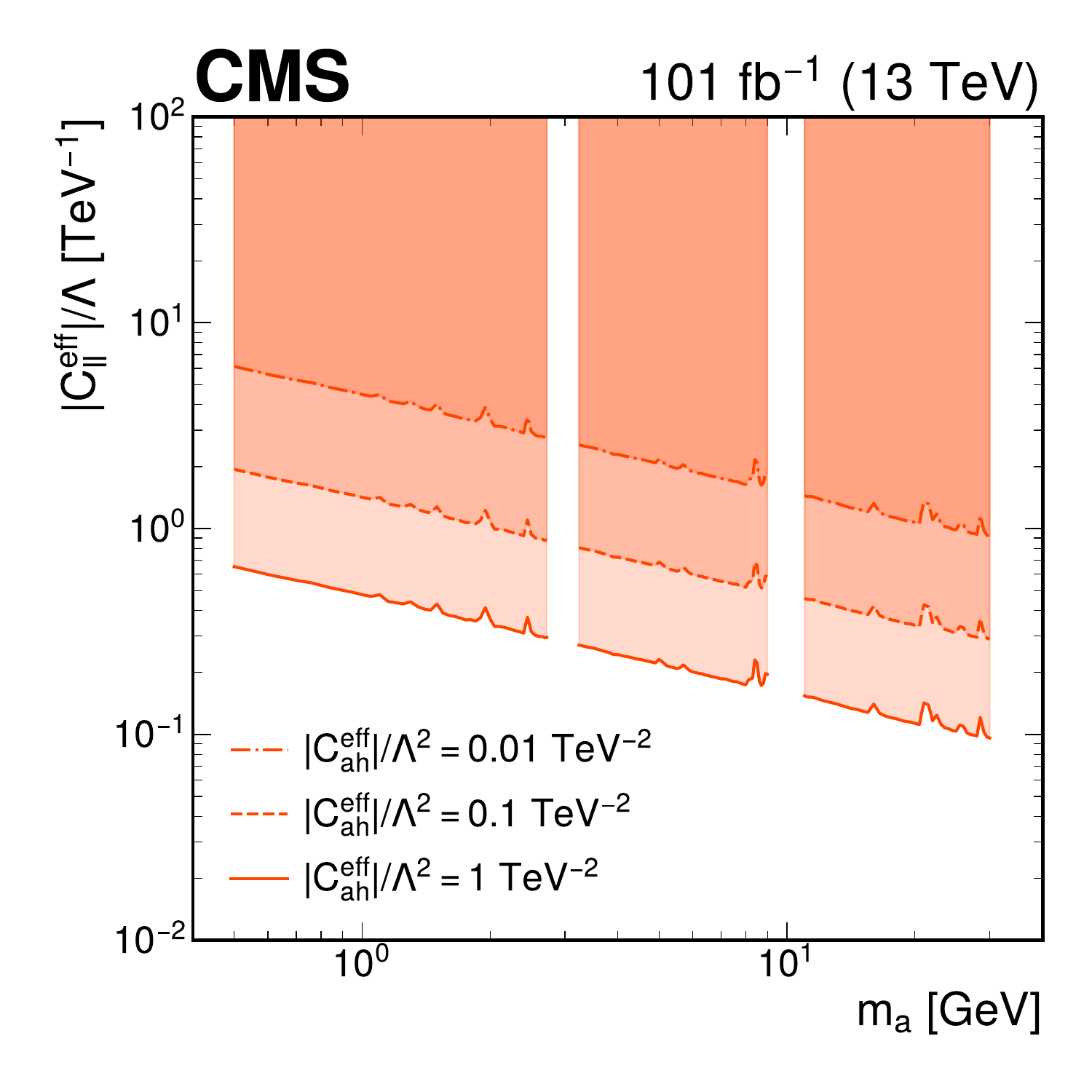}
\caption{Left: The 95\% \CL observed upper limits on the effective coupling \CahLambda of the ALP to the SM Higgs assuming the branching fraction of the ALP to muons is 1 (blue) and 0.1 (orange), for both the expected (dashed) and observed (solid) limits. Right: The 95\% \CL observed upper limits on the effective coupling $\cll/\Lambda$ of the ALP to the SM leptons. The orange shaded region represents the parameter space excluded by this search under three choices of \CahLambda: $1\TeV^{-2}$ (solid), $0.1\TeV^{-2}$ (dashed), and $0.01\TeV^{-2}$ (dotted). 
}
\label{fig:ALPLimit}
\end{figure}

For the ALP scenario, the model-independent limit in Fig.~\ref{fig:ModelIndependentLimit} can be used to set the limit on the ratio of the effective coupling between the ALP and the SM Higgs boson, $C^{\text{eff}}_{\Pa\Ph}$, and the square of the new-physics energy scale, $\Lambda$, or \CahLambda. The partial width of the Higgs boson decay to two ALPs can be found in Ref.~\cite{Bauer:2017ris}. We assume the Higgs boson mass is 125\GeV, the partial width of Higgs boson decay to SM particles is 4.1\MeV, the production cross section of the SM \PH via \Pggf fusion at the LHC is 48.93\unit{pb} at a center-of-mass energy of 13\TeV~\cite{Cepeda:2019klc}, the vacuum expectation value of the Higgs field is 246\GeV, and the branching fraction of the ALP to muons is 100\% (or 10\%). The observed 95\% \CL upper limit on \CahLambda as a function of the ALP mass \MPa is shown in Fig.~\ref{fig:ALPLimit} (left). Through this analysis, we achieve further reach and greater sensitivity than Run 1 LHC searches~\cite{ATLAS:2015unc, CMS:2015twz, CMS:2017dmg} (summarized in Ref.~\cite{Bauer:2017ris}). Similarly, with different choices of \CahLambda ($1\TeV^{-2}$, $0.1\TeV^{-2}$ and $0.01\TeV^{-2}$), and a decay width of the ALP to leptons as presented in Ref.~\cite{Bauer:2017ris}, we set limits on the ratio of the effective coupling between the ALP and SM leptons, \cll, to the new-physics energy scale, or $\cll/\Lambda$, as shown in Fig.~\ref{fig:ALPLimit} (right). The limit on $\cll/\Lambda$ depends directly on the value assumed for the total ALP width. We fix the ALP total width to a constant reference point of $10^{-5}\GeV$ for all mass points.

\begin{figure}[h!]
\centering
\includegraphics[width=0.75\linewidth]{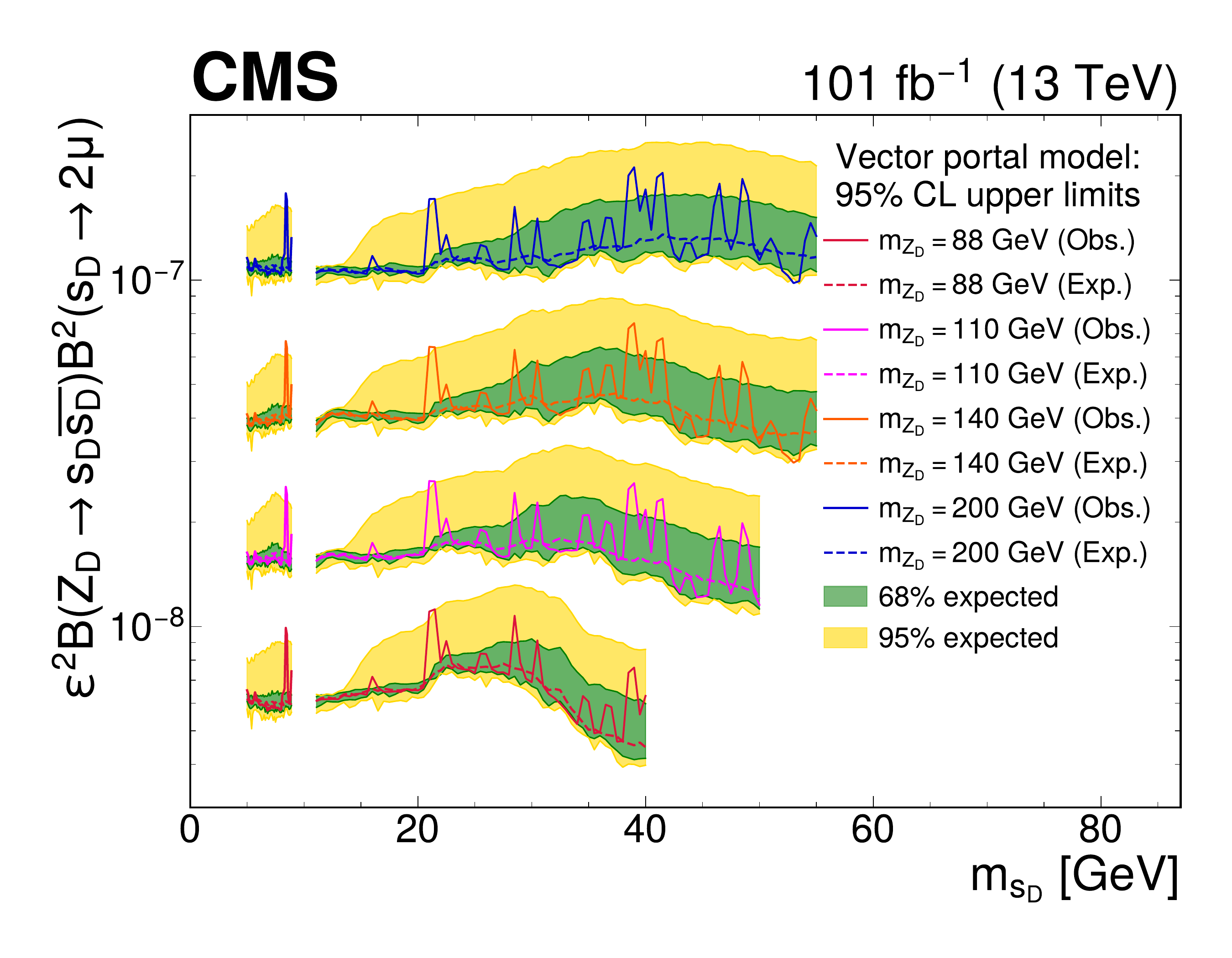}
\caption{The 95\% \CL observed upper limits on $\varepsilon^{2}\mathcal{B}(\ZDark \to \sDark \AsDark)\mathcal{B}^{2}(\sDark \to 2\PGm)$ for the vector portal model as a function of the dark scalar mass \MsDark and dark vector boson mass \MZDark.}
\label{fig:VectorPortalLimitEp}
\end{figure}

{\tolerance=1600
For the vector portal benchmark model, we set a 95\% \CL upper limit on the product of the production cross section of the \ZDark, the branching fraction of the decay of \ZDark to \sDark and $\AsDark$, and the squared branching fraction of the decay of \sDark to two oppositely charged muons: \mbox{$\sigma(\PP \to \ZDark)\mathcal{B}(\ZDark \to \sDark \AsDark)\mathcal{B}^{2}(\sDark \to 2\PGm)$} as a function of the dark scalar mass \MsDark and the dark vector boson mass \MZDark. This limit is subsequently translated to the limit on the product of the square of the kinetic mixing parameter, $\varepsilon^{2}$, times the branching fraction of the decay of \ZDark to a pair of dark scalar bosons \sDark, times the squared branching fraction of the decay of the $s_{D}$ to a dimuon: $\varepsilon^{2}\mathcal{B}(\ZDark \to \sDark \AsDark)\mathcal{B}^{2}(\sDark \to 2\PGm)$. As the $s_{D}$ may also decay into SM particles other than dimuons, the branching fraction of its decay into dimuons must be taken into account in this limit, for various \MsDark and \MZDark, which is shown in Fig.~\ref{fig:VectorPortalLimitEp} for various \MsDark and \MZDark.
\par}

\begin{figure}[h!]
\centering
\includegraphics[width=\linewidth]{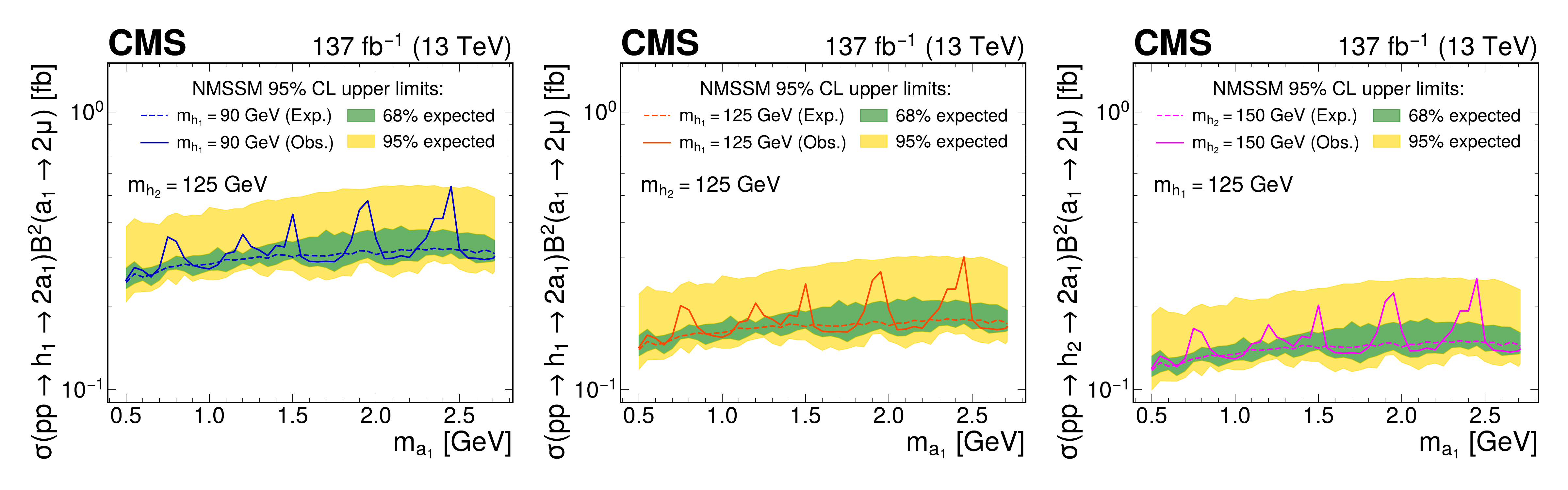}
\caption{The 95\% \CL observed upper limits for $\sigma(\PP \to \PhOneTwo \to 2\PaOne)\mathcal{B}^2(\PaOne \to 2\PGm)$ for the NMSSM as a function of \MPaOne for three choices of \MPhOneTwo. The left and middle plots have \MPhTwo fixed to 125\GeV and the right plot has \MPhOne fixed to 125\GeV. The data presented here reflect the results of the 2017 and 2018 data sets combined with the previously published results of the 2016 data set in \cite{CMS:2018jid}.}
\label{fig:NMSSMLimit}
\end{figure}

{\tolerance=10000
For the NMSSM scenario, the 95\% \CL observed upper limits are derived for \mbox{$\sigma(\PP \to \PhOneTwo \to 2\PaOne)\mathcal{B}^2(\PaOne\to 2\PGm)$} as a function of \MPaOne with $\MPhOne = 90\GeV$ \mbox{(Fig.~\ref{fig:NMSSMLimit} left)} and $\MPhOne = 125\GeV$ \mbox{(Fig.~\ref{fig:NMSSMLimit} middle)}, with \MPhTwo fixed at 125\GeV; and $\MPhTwo = 150\GeV$ \mbox{(Fig.~\ref{fig:NMSSMLimit} right)} with \MPhOne fixed at 125\GeV. It is assumed that all contributions come from either \PhOne or $\PhTwo$; there is no case in which both \PhOne and \PhTwo decay to the \PaOne. The branching fraction of the \PaOne decay to a dimuon is dependent on the \PaOne mass~\cite{Dermisek:2010mg}.
\par}

\begin{figure}[h!]
\centering
\includegraphics[width=0.8\linewidth]{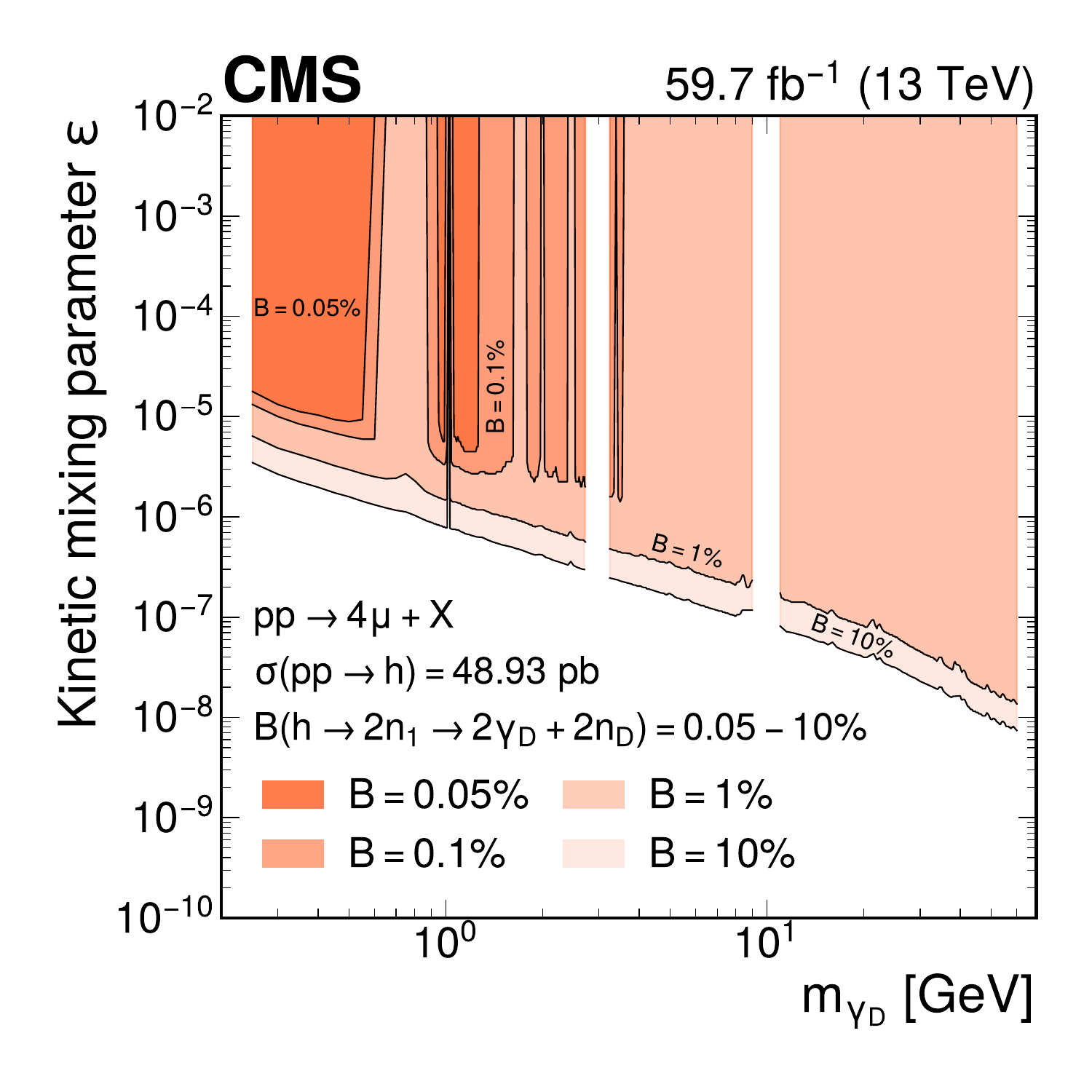}
\caption{The 95\% \CL observed upper limits (black solid curves) from this search as interpreted in the dark SUSY scenario for the process $\PP \to \Ph \to 2\nOne \to 2\gammaDark + 2\nDark \to 4\PGm + \PX$, with $\MnOne = 60\GeV$ and $\MnDark = 1\GeV$. The limits are presented in the plane of $\varepsilon$ and \MgammaDark. The color gradient represents different branching fraction assumptions for \mbox{$\mathcal{B}(\Ph \to 2\nOne \to 2\gammaDark + 2\nDark)$}, ranging from dark orange (0.05\%) to light orange (10\%). The degradation of the limit around 1\GeV is attributed to the drop in the dimuon branching fraction $\mathcal{B}(\gammaDark \to 2\mu)$ due to the dimuon resonance of the $\PGf$ meson~\cite{Batell:2009yf}.}
\label{fig:DarkSUSYLimit}
\end{figure}

For the dark SUSY scenario, a 95\% \CL observed upper limit is set on the product of the \Ph production cross section and the branching fractions of the \Ph decay to a pair of dark photons. The limit set by this experimental search is presented in Fig.~\ref{fig:DarkSUSYLimit} as areas excluded in a two-dimensional plane of $\varepsilon$ and \MgammaDark. Here, the Higgs boson mass is taken to be 125\GeV, and the production cross section of the Higgs boson via gg fusion at the LHC is 48.93\unit{pb} at 13\TeV~\cite{Cepeda:2019klc}.

For this search, limits are shown for the branching fraction of the \Ph decay to two dark photons in the range 0.05--10\%. The kinetic mixing parameter, $\varepsilon$, the mass of the dark photon \MgammaDark, the branching fraction of the \gammaDark decays to a dimuon, and the proper decay length of the dark photon \cTgammaDark are related via an analytic function that is solely dependent on the dark photon mass~\cite{Batell:2009yf}. The proper decay length of the dark photon is allowed to vary from 0 to 100\mm and \MgammaDark can range from 0.25 to 58\GeV. With the extended ranges in dimuon mass (0.21--60\GeV), this analysis excludes a previously unexplored area in the parameter space of $\varepsilon$ and \MgammaDark compared to the previous model-independent \mbox{search (0.25--8.5\GeV)~\cite{CMS:2018jid}}.

\clearpage
\section{Summary}\label{sec:summary}
A search for pairs of new bosons \Pa that subsequently decay to pairs of oppositely charged muons is presented. This search is performed using data samples collected by the CMS experiment in 2017 and 2018, corresponding to 41.5\fbinv for the prompt signal in 2017 and 59.7\fbinv for both the prompt and displaced signals in 2018, for proton-proton collisions at $\sqrt{s} = 13\TeV$. The results are also combined with a similar analysis performed by the CMS Collaboration~\cite{CMS:2018jid}, which analyzed a smaller data set collected in 2016 corresponding to 35.9\fbinv of proton-proton collisions at $\sqrt{s} = 13\TeV$.
Additionally, both the mass range of the boson \Pa and the maximum possible displacement of its decay vertex are extended compared to the previous version of this analysis. The distribution of events in the signal region is consistent with standard model expectations. Model-independent 95\% confidence level upper limits on the product of the production cross section of pairs of new bosons, the square of the branching fraction to dimuons, and the acceptance are set over the mass range $0.21 < \MPa < 60\GeV$ and are found to vary between 0.049 and 0.247\unit{fb}. These model-independent limits are then interpreted in the context of an axion-like particle model, a vector portal model, a next-to-minimal supersymmetric standard model, and a dark supersymmetry scenario with non-negligible boson proper decay length of up to $\ctau= 100\mm$.

\begin{acknowledgments}
    We congratulate our colleagues in the CERN accelerator departments for the excellent performance of the LHC and thank the technical and administrative staffs at CERN and at other CMS institutes for their contributions to the success of the CMS effort. In addition, we gratefully acknowledge the computing centers and personnel of the Worldwide LHC Computing Grid and other centers for delivering so effectively the computing infrastructure essential to our analyses. Finally, we acknowledge the enduring support for the construction and operation of the LHC, the CMS detector, and the supporting computing infrastructure provided by the following funding agencies: SC (Armenia), BMBWF and FWF (Austria); FNRS and FWO (Belgium); CNPq, CAPES, FAPERJ, FAPERGS, and FAPESP (Brazil); MES and BNSF (Bulgaria); CERN; CAS, MoST, and NSFC (China); MINCIENCIAS (Colombia); MSES and CSF (Croatia); RIF (Cyprus); SENESCYT (Ecuador); ERC PRG, RVTT3 and MoER TK202 (Estonia); Academy of Finland, MEC, and HIP (Finland); CEA and CNRS/IN2P3 (France); SRNSF (Georgia); BMBF, DFG, and HGF (Germany); GSRI (Greece); NKFIH (Hungary); DAE and DST (India); IPM (Iran); SFI (Ireland); INFN (Italy); MSIP and NRF (Republic of Korea); MES (Latvia); LMTLT (Lithuania); MOE and UM (Malaysia); BUAP, CINVESTAV, CONACYT, LNS, SEP, and UASLP-FAI (Mexico); MOS (Montenegro); MBIE (New Zealand); PAEC (Pakistan); MES and NSC (Poland); FCT (Portugal); MESTD (Serbia); MCIN/AEI and PCTI (Spain); MOSTR (Sri Lanka); Swiss Funding Agencies (Switzerland); MST (Taipei); MHESI and NSTDA (Thailand); TUBITAK and TENMAK (Turkey); NASU (Ukraine); STFC (United Kingdom); DOE and NSF (USA).
      
    \hyphenation{Rachada-pisek} Individuals have received support from the Marie-Curie program and the European Research Council and Horizon 2020 Grant, contract Nos.\ 675440, 724704, 752730, 758316, 765710, 824093, 101115353, 101002207, and COST Action CA16108 (European Union); the Leventis Foundation; the Alfred P.\ Sloan Foundation; the Alexander von Humboldt Foundation; the Science Committee, project no. 22rl-037 (Armenia); the Belgian Federal Science Policy Office; the Fonds pour la Formation \`a la Recherche dans l'Industrie et dans l'Agriculture (FRIA-Belgium); the F.R.S.-FNRS and FWO (Belgium) under the ``Excellence of Science -- EOS" -- be.h project n.\ 30820817; the Beijing Municipal Science \& Technology Commission, No. Z191100007219010 and Fundamental Research Funds for the Central Universities (China); the Ministry of Education, Youth and Sports (MEYS) of the Czech Republic; the Shota Rustaveli National Science Foundation, grant FR-22-985 (Georgia); the Deutsche Forschungsgemeinschaft (DFG), among others, under Germany's Excellence Strategy -- EXC 2121 ``Quantum Universe" -- 390833306, and under project number 400140256 - GRK2497; the Hellenic Foundation for Research and Innovation (HFRI), Project Number 2288 (Greece); the Hungarian Academy of Sciences, the New National Excellence Program - \'UNKP, the NKFIH research grants K 131991, K 133046, K 138136, K 143460, K 143477, K 146913, K 146914, K 147048, 2020-2.2.1-ED-2021-00181, and TKP2021-NKTA-64 (Hungary); the Council of Science and Industrial Research, India; ICSC -- National Research Center for High Performance Computing, Big Data and Quantum Computing and FAIR -- Future Artificial Intelligence Research, funded by the NextGenerationEU program (Italy); the Latvian Council of Science; the Ministry of Education and Science, project no. 2022/WK/14, and the National Science Center, contracts Opus 2021/41/B/ST2/01369 and 2021/43/B/ST2/01552 (Poland); the Funda\c{c}\~ao para a Ci\^encia e a Tecnologia, grant CEECIND/01334/2018 (Portugal); the National Priorities Research Program by Qatar National Research Fund;  MCIN/AEI/10.13039/501100011033, ERDF ``a way of making Europe", and the Programa Estatal de Fomento de la Investigaci{\'o}n Cient{\'i}fica y T{\'e}cnica de Excelencia Mar\'{\i}a de Maeztu, grant MDM-2017-0765 and Programa Severo Ochoa del Principado de Asturias (Spain); the Chulalongkorn Academic into Its 2nd Century Project Advancement Project, and the National Science, Research and Innovation Fund via the Program Management Unit for Human Resources \& Institutional Development, Research and Innovation, grant B39G670016 (Thailand); the Kavli Foundation; the Nvidia Corporation; the SuperMicro Corporation; the Welch Foundation, contract C-1845; and the Weston Havens Foundation (USA).
\end{acknowledgments}

\bibliography{auto_generated}

\cleardoublepage \appendix\section{The CMS Collaboration \label{app:collab}}\begin{sloppypar}\hyphenpenalty=5000\widowpenalty=500\clubpenalty=5000
\cmsinstitute{Yerevan Physics Institute, Yerevan, Armenia}
{\tolerance=6000
A.~Hayrapetyan, A.~Tumasyan\cmsAuthorMark{1}\cmsorcid{0009-0000-0684-6742}
\par}
\cmsinstitute{Institut f\"{u}r Hochenergiephysik, Vienna, Austria}
{\tolerance=6000
W.~Adam\cmsorcid{0000-0001-9099-4341}, J.W.~Andrejkovic, T.~Bergauer\cmsorcid{0000-0002-5786-0293}, S.~Chatterjee\cmsorcid{0000-0003-2660-0349}, K.~Damanakis\cmsorcid{0000-0001-5389-2872}, M.~Dragicevic\cmsorcid{0000-0003-1967-6783}, P.S.~Hussain\cmsorcid{0000-0002-4825-5278}, M.~Jeitler\cmsAuthorMark{2}\cmsorcid{0000-0002-5141-9560}, N.~Krammer\cmsorcid{0000-0002-0548-0985}, A.~Li\cmsorcid{0000-0002-4547-116X}, D.~Liko\cmsorcid{0000-0002-3380-473X}, I.~Mikulec\cmsorcid{0000-0003-0385-2746}, J.~Schieck\cmsAuthorMark{2}\cmsorcid{0000-0002-1058-8093}, R.~Sch\"{o}fbeck\cmsorcid{0000-0002-2332-8784}, D.~Schwarz\cmsorcid{0000-0002-3821-7331}, M.~Sonawane\cmsorcid{0000-0003-0510-7010}, S.~Templ\cmsorcid{0000-0003-3137-5692}, W.~Waltenberger\cmsorcid{0000-0002-6215-7228}, C.-E.~Wulz\cmsAuthorMark{2}\cmsorcid{0000-0001-9226-5812}
% ...
\par}
\vskip\cmsinstskip
\dag:~Deceased\\
$^{1}$Also at Yerevan State University, Yerevan, Armenia\\
$^{2}$Also at TU Wien, Vienna, Austria\\
% ...
\end{sloppypar}
\end{document}